\documentclass[a4paper,twocolumn,11pt,unpublished,accepted=2026-06-03]{quantumarticle}
\pdfoutput=1
\usepackage[utf8]{inputenc}
\usepackage[english]{babel}
\usepackage[T1]{fontenc}
\usepackage{amsmath}
\usepackage{hyperref}

\usepackage[numbers,sort&compress]{natbib}

\usepackage{bbold}

\usepackage{tikz}
\usepackage{lipsum}

\usepackage{placeins}

\usepackage{mathtools}
\usepackage{xfrac}
\usepackage{enumitem}
\usepackage{xcolor}
\usepackage{mathrsfs}
\usepackage{bm}
\usepackage{amssymb}
\usepackage{siunitx}
\usepackage{tabularray}
\UseTblrLibrary{booktabs}
\UseTblrLibrary{siunitx}
\usepackage{xspace}
\usetikzlibrary{decorations.pathreplacing}
\usetikzlibrary{arrows.meta}
\usepackage{csquotes}
\usepackage{placeins}
\usepackage[numbers]{natbib}
\DeclarePairedDelimiter{\bra}{\langle}{\rvert}
\DeclarePairedDelimiter{\ket}{\lvert}{\rangle}
\DeclarePairedDelimiterX{\braket}[2]{\langle}{\rangle}{
#1 \delimsize| #2
}

\newcommand*\diff{\mathop{}\!\mathrm{d}}
\newcommand*\Diff[1]{\mathop{}\!\mathrm{d}^#1}
\DeclarePairedDelimiterX{\comm}[2]{[}{]}{
  \hspace{0.1em} #1 \delimsize, #2 \hspace{0.1em} 
}
\newcommand*\var{\ensuremath{\hat{H}_\text{A}^\text{Var}(\bm{g})}\xspace}
\newcommand*\varwo{\ensuremath{\hat{H}_\text{A}^\text{Var}}\xspace}
\newcommand*\varopt{\ensuremath{\hat{H}_\text{A}^\text{Var}(\bm{g}^\text{opt})}\xspace}
\DeclareMathOperator{\Tr}{Tr}
\newcommand*\gopt{\ensuremath{\bm{g}^\text{opt}}\xspace}

\newcommand*\obs{\ensuremath{\hat{\mathcal{O}}^\text{A}}}

\newcommand*\e{\ensuremath{\mathrm{e}}}
\newcommand*\BW{\ensuremath{\hat{H}_\text{A}^\text{BW}}\xspace}
\newcommand*\BWV{\ensuremath{\hat{H}_\text{A}^\text{BWV}}\xspace}
\newcommand*\hi{\ensuremath{\hat{h}_i}\xspace}
\newcommand*\NA{\ensuremath{N_\text{A}}\xspace}
\newcommand*\corr{\ensuremath{\hat{H}_\text{A}^{\text{c}}}\xspace}
\newcommand*\TMAX{\ensuremath{T_\text{max}}\xspace}
\newcommand*\dt{\ensuremath{\mathrm{\Delta} t}\xspace}
\newcommand*\rmax{\ensuremath{r_\text{max}}\xspace}
\newcommand*\copt{\ensuremath{\mathcal{C}(\gopt)}\xspace}
\newcommand*\increment{\ensuremath{\mathrm{\Delta}}\xspace}

\definecolor{wong_sky_blue} {RGB}{86, 180, 233}
\definecolor{wong_blue} {RGB} {0, 114, 178}
\definecolor{wong_orange} {RGB} {230, 159, 0}

\begin{document}

\title{Exploring Variational Entanglement Hamiltonians}

\author{Yanick S. Kind}
\affiliation{Condensed Matter Theory, TU Dortmund University,
Otto-Hahn-Straße 4, 44227 Dortmund, Germany}
\affiliation{Institute of Software Technology, German Aerospace Center (DLR), 51147 Cologne, Germany}
\orcid{0009-0007-1710-3507}

\author{Benedikt Fauseweh}
\affiliation{Condensed Matter Theory, TU Dortmund University,
Otto-Hahn-Straße 4, 44227 Dortmund, Germany}
\affiliation{Institute of Software Technology, German Aerospace Center (DLR), 51147 Cologne, Germany}
\email{benedikt.fauseweh@tu-dortmund.de}
\orcid{0000-0002-4861-7101}

\begin{abstract}
Recent advances in analog and digital quantum-simulation platforms have enabled exploration of the spectrum of entanglement Hamiltonians via variational algorithms. In this work we analyze the convergence properties of the variationally obtained solutions and compare them to numerically exact calculations in quantum critical systems. We demonstrate that interpreting the cost functional as an integral permits the deployment of iterative quadrature schemes, thereby reducing the required number of measurements by 
more than an order of magnitude even in the presence of noise. We further show that a modified ansatz captures deviations from the Bisognano–Wichmann form in lattice models, improves convergence, improves trainability and provides a cost-function-level diagnostic for quantum phase transitions. Finally, we establish that a low cost value does not by itself guarantee convergence in trace distance. Nevertheless, it faithfully reproduces degeneracies and spectral gaps, which are essential for applications to topological phases.
\end{abstract}

Generating and controlling novel quantum states is a central objective in condensed-matter physics and in the development of emerging quantum devices. Complex many-body states are often highly entangled, rendering their analysis by classical numerical methods infeasible. Consequently, experiments that prepare such states frequently provide the only viable means of investigation—an idea at the heart of quantum simulation \cite{Feynman1982,Lloyd1073,Daley2022,Fauseweh2024}. By tuning control parameters and performing measurements we can explore systems beyond classical computational limits, with notable examples in trapped-ion and ultracold-atom platforms \cite{RevModPhys.93.025001,Eisert2015,Zhang2017,PhysRevX.12.011018,Bernien2017,doi:10.1126/science.aal3837,Bloch2012,Blatt2012,Houck2012,Aspuru-Guzik2012,doi:10.1126/science.abi8794}.

Variational quantum algorithms (VQAs) \cite{Cerezo2021} have emerged as a valuable tool in quantum simulation. These hybrid quantum–classical schemes leverage the complementary strengths of a classical computer to efficiently optimize scalar objective functions and of a quantum processor to represent, manipulate, and measure states in high-dimensional Hilbert spaces. Potential applications include the preparation of ground \cite{Peruzzo2014,PhysRevX.6.031007,Kandala2017,PhysRevResearch.6.043254}, excited \cite{Higgott2019variationalquantum,PhysRevResearch.1.033062,Atas2021}, time-evolved \cite{Gibbs2022,PhysRevResearch.4.023097,Barison2021efficientquantum}, and Floquet states \cite{Fauseweh2023quantumcomputing,kumar2025floquetadaptvqequantumalgorithmsimulate}.

Recently, Kokail \textit{et al.} proposed a variational algorithm to learn the entanglement Hamiltonian of a given ground state \cite{PhysRevLett.127.170501,Zache2022entanglement} and demonstrated it experimentally in trapped-ion systems \cite{Kokail2021,Joshi2023}. Ultracold atomic gases have also been used to variationally determine the entanglement Hamiltonian \cite{Redon2024}. Combined with entanglement spectroscopy, the learned Hamiltonian enables determination of the low-lying entanglement spectrum without exponential resource costs. Since the seminal work by Li and Haldane \cite{PhysRevLett.101.010504}, the entanglement spectrum has become an important diagnostic not only for topologically ordered phases but also for symmetry-protected topological phases \cite{PhysRevLett.104.130502,PhysRevB.81.064439}, tensor-network approaches \cite{PhysRevLett.132.086503},  many-body-localized systems \cite{PhysRevLett.117.160601} and to detect quantum phase transitions \cite{PhysRevLett.109.237208, PhysRevB.87.235107, PhysRevB.106.014306}. Notably the measurement of the entanglement spectrum of a symmetry protected topological state was also realized on IBM quantum computers \cite{PhysRevLett.121.086808}.

The variational algorithm employs an ansatz based on the Bisognano–Wichmann (BW) form of the entanglement Hamiltonian \cite{10.1063/1.522605,10.1063/1.522898,Dalmonte2018}. While this form is exact for ground states of relativistic quantum field theories (QFTs), it is generally only an approximation on a lattice. As a result, resolving spectral gaps and degeneracies with high fidelity can be demanding. Moreover, optimization of the cost function entails time evolution under the ansatz Hamiltonian and measurement of observables at multiple time points, leading to substantial measurement overhead.

In this paper we address these challenges by analyzing the convergence properties of the variational algorithm. We perform classical simulations for several one-dimensional spin models, contrast the simple right point integration rule \cite{Kokail2021, PhysRevLett.127.170501} with more sophisticated quadrature schemes, and demonstrate the superiority of the latter, even in the presence of noise. We further show that a BW violating ansatz can improve trainability and thereby lower the measurement costs of optimization and render the solution insensitive to observation time. Finally, we establish that a small cost value does not necessarily imply high fidelity as measured by the trace distance; nevertheless, it reliably indicates spectral gaps and degeneracies in the entanglement spectrum, as we demonstrate for the symmetry-protected Haldane phase in a $S=1$ chain.

\section{Theoretical foundation}
\subsection{Schmidt decomposition and entanglement Hamiltonian}
Given a composite Hilbert space $\mathscr{H} = \mathscr{H}_\text{A} \otimes \mathscr{H}_\text{B}$, composed of two subsystems A and B with their respective 
Hilbert spaces $\mathscr{H}_\text{A}$ and $\mathscr{H}_\text{B}$ of dimensions $d_\text{A} = \text{dim}(\mathscr{H}_\text{A})$
and $d_\text{B} = \text{dim}(\mathscr{H}_\text{B})$, spanned by the orthonormal bases $\{ \ket{\mu_\text{A}^i} \}$ and $\{ \ket{\mu_\text{B}^i} \}$,
a general pure state $\ket{\Psi} \in \mathscr{H}$ can be written as 
\begin{equation}
    \ket{\Psi} = \sum_{i = 1}^{d_\text{A}} \sum_{j = 1}^{d_\text{B}} M_{ij} \ket{\mu_\text{A}^i} \otimes \ket{\mu_\text{B}^j}, \label{eqn:GenBiPsi}
\end{equation}
where the rank $\chi \leq \text{min}(d_\text{A}, d_\text{B})$ of the complex matrix $M$ is called the Schmidt rank \cite{RevModPhys.81.865}.
The entanglement matrix $M$ can be brought into a diagonal form $D$ via 
a singular value decomposition (SVD) \cite{10.1093/acprof:oso/9780198785781.003.0004}
\begin{equation}
    M = UDV^{\dagger}.
\end{equation}
The matrices $U$ and $V$ are of size $d_\text{A} \times \text{min}(d_\text{A}, d_\text{B})$ and 
$d_\text{B} \times \text{min}(d_\text{A}, d_\text{B})$, respectively, and obey $U^{\dagger}U = \mathbb{1}$ and $VV^{\dagger} = \mathbb{1}$.
The non-negative entries (the singular values of $M$) of the diagonal matrix $D$ with dimension $\text{min}(d_\text{A},d_\text{B})$ are called Schmidt-coefficients \cite{Nielsen_Chuang_2010}  
and can be expressed as $\mathrm{e}^{-\sfrac{\xi_\alpha}{2}}$.
Using the SVD, Eq. \eqref{eqn:GenBiPsi} reads
\begin{equation}
    \ket{\Psi} = \sum_{i=1}^{d_\text{A}} \sum_{j=1}^{d_\text{B}} \sum_{\alpha=1}^{\text{min}(d_\text{A},d_\text{B})} \e^{-\sfrac{\xi_\alpha}{2}} U_{i\alpha} V^*_{j\alpha} \ket{\mu_\text{A}^i} \otimes \ket{\mu_\text{B}^j}.
\end{equation}
Defining a new orthonormal basis set $\{ \ket{\Phi^{\alpha}_\text{A}} = \sum_{i=1}^{d_\text{A}} U_{i \alpha} \ket{\mu_\text{A}^i} \}$ and 
$\{ \ket{\Phi^{\alpha}_\text{B}} = \sum_{j=1}^{d_\text{B}} V^*_{j \alpha} \ket{\mu_\text{B}^j} \}$ yields
\begin{equation}
    \ket{\Psi} = \sum_{\alpha = 1}^{\chi} \e^{-\xi_{\alpha}/2} \ket{\Phi^{\alpha}_\text{A}} \otimes \ket{\Phi^{\alpha}_\text{B}}, \label{eqn:schmidt}
\end{equation}
where $\{ \xi_{\alpha} \}$ will be referred to as the Entanglement Spectrum (ES).
Since the rank is preserved under a SVD, the number of non-zero singular values coincides with the Schmidt rank $\chi$ \cite{lay2015}, and thus, the sum in Eq. \eqref{eqn:schmidt} is restricted to $\chi$.
The lower and upper bound of summation will be dropped from now on as long as it is unambiguous. 
The reduced density matrix of the state $\hat{\rho} = \ket{\Psi}\bra{\Psi}$ on subsystem A  after tracing out the degrees of freedom related to subsystem B is given by
\begin{equation}
    \hat{\rho}_\text{A} = \Tr_\text{B} \left [ \hat{\rho} \right ] = \sum_{\alpha} \e^{-\xi_{\alpha}} \ket{\Phi_\text{A}^{\alpha}}\bra{\Phi_\text{A}^{\alpha}} 
   =\mathrel{\mathop:} \e^{-\hat{H}_\text{A}},
\end{equation}  
which defines the Entanglement Hamiltonian (EH) $\hat{H}_\text{A}$ \cite{PhysRevLett.127.170501,Rottoli_2024}.
The EH and its non-negative eigenvalues $\{ \xi_{\alpha} \} $, the ES, completely characterize all correlations in partition A \cite{PhysRevLett.127.170501}
and reveal much more than the entanglement entropy \cite{PhysRevLett.101.010504} or the entanglement witness \cite{PhysRevB.106.014306}.
In general, it is hard to derive an analytical form of the EH especially for lattice theories.
The BW theorem (Section \ref{sec:BW}) allows one to obtain the EH analytically
for specific cases for QFTs.
\subsection{Bisognano-Wichmann theorem}
\label{sec:BW}
In a $d+1$-dimensional Poincaré invariant QFT with a local Hamiltonian density $\hat{\mathcal{H}}(\bm{x})$, the \textit{exact} EH of the ground state for the special case of a bipartition of an infinite system A (A $= \{ \bm{x} \in \mathbb{R}^d | x_1 > 0\}$) 
is
\begin{equation}
    \hat{H}_\text{A} = \int_\text{A} \Diff{d}x \, \beta(x_1) \hat{\mathcal{H}}(\bm{x}) + c'\label{eqn:BW}
\end{equation}
with $\beta(x_1) = \frac{2\pi}{c} x_1$  \cite{https://doi.org/10.1002/andp.202200064, PhysRevB.98.134403, PhysRevLett.127.170501}, 
whereby the speed of sound $c$ of the underlying QFT is set to unity from now on.
The constant $c'$ ensures the normalization $\Tr[\hat{\rho}_\text{A}] = 1$.
This is the seminal BW-theorem.
In Eq. \eqref{eqn:BW} it becomes apparent that the EH is a deformation of the system Hamiltonian \cite{PhysRevLett.127.170501}.
Additionally, the reduced density matrix $\hat{\rho}_\text{A}$ can be interpreted as a thermal state with a locally varying entanglement temperature,
which is very high near the entanglement cut (boundary between both partitions) and decreases with $\sfrac{1}{x_1}$ away from it \cite{PhysRevLett.127.170501}, see also Fig.~\ref{fig:BW}.\\
\begin{figure}
    \centering
     \resizebox{\linewidth}{!}{
    \pgfdeclarehorizontalshading{BW}{100bp}
 {color(0bp)=(wong_orange);  color(100bp)=(wong_sky_blue)}

 \begin{tikzpicture}[shading = BW]
    \shade (3.5,0) rectangle node[black] {\large \textsc{A}} (7,2.5);
    \draw[thick,->] (3.5,2.9) -- (7,2.9) node[midway, above] {\large $x_1$};
    \node at (8.5,1.25) {\Large $\beta \propto x_1$};
    \draw (0,0) -- (3.5,0);
    \draw (3.5,0) -- (3.5, 2.5);
    \draw (0,2.5) -- (3.5, 2.5);
    \draw[dashed] (0, 0) -- (0,2.5);
    \draw (3.5,0) -- (7,0);
    \draw[dashed]  (7,0) -- (7, 2.5);
    \draw (3.5, 2.5) -- (7,2.5); 
    \node at (6,2) {\large $\hat{\rho}_\text{A} = \e^{-\hat{H}_\text{A}}$};
    \node at (1.75, 1.25) {\large\textsc{B}};
\end{tikzpicture}
    }
    \caption{Interpretation of the reduced density matrix $\hat{\rho}_\text{A}$ as a thermal state with a locally varying temperature, the entanglement temperature.
    The inverse entanglement temperature takes the form of a linear ramp, and thus, the entanglement temperature decreases as $\propto \sfrac{1}{x_1}$, indicated by 
    the color gradient from orange to blue.}
    \label{fig:BW}
\end{figure}
For lattice systems, it is straightforward to propose a discretized version of Eq. \eqref{eqn:BW} s.t.
\begin{equation}
    \hat{H}_\text{A} \approx \sum_{i\in\text{A}} g_i \hat{h}_i \label{eqn:discrBW} + c',
\end{equation}
where the substitution $\beta(x_1) \to g_i$ and $\hat{\mathcal{H}}(\bm{x}) \to \hat{h}_i$ with $\hat{h}_i$ as a quasi-local few-body operator for the $i$-th lattice site is utilized. 
A natural question is whether the BW theorem works for lattice systems since it is defined for relativistic QFTs at first.
Although the presence of a lattice breaks the Lorentz invariance \cite{https://doi.org/10.1002/andp.202200064} (even when it 
is recovered as a low-energy symmetry\cite{PhysRevB.98.134403}), analytical and numerical calculations \cite{https://doi.org/10.1002/andp.202200064,PhysRevB.98.134403,PhysRevB.100.155122,Rottoli_2024}
suggest that the discretized version of the BW theorem \eqref{eqn:discrBW} is often a good first approximation for lattice systems. 
\subsection{Conformal extensions}
\label{subsec:Confext}
For systems, which have conformal symmetry in addition to Lorentz invariance, the BW theorem (Eq. \eqref{eqn:BW}) can be extended
to different geometries \cite{PhysRevB.98.134403}. 
In this work we focus one-dimensional systems.
\begin{figure}
    \centering
    \resizebox{\linewidth}{!}{
    \pgfdeclareverticalshading{CFT_periodic}{100bp}
 {color(0bp)=(wong_orange);  color(20bp)=(wong_orange); color(50bp)=(wong_sky_blue); color(80bp)=(wong_orange); color(100bp)=(wong_orange)}
 \pgfdeclarehorizontalshading{CFT_sin}{100bp}
 {color(0bp)=(wong_orange);  color(20bp)=(wong_orange); color(100bp)=(wong_sky_blue)}

 \begin{tikzpicture}[]
    \shade[shading=CFT_periodic,shading angle=35] (110:0.9cm) arc [start angle = 110, end angle=-40, radius=0.9cm] --++(-40:0.55cm) arc [start angle=-40, end angle=110, radius=1.45cm] --++(110:-0.55cm)--cycle;  
    %\draw[shading=CFT_periodic,shading angle=15]
    %      (110:1)
    %  arc (110:-40:1)
    %  --  (-40:0.35)
    %  arc (-40:-110:1.35)
    %  -- cycle;
    \draw circle (1.45);     
    \draw circle (0.9); 
    \draw (110:0.9cm) -- (110:1.45cm);
    \draw (-40:0.9cm) -- (-40:1.45cm);
    \node at (35:1.21) {\LARGE A};
    \node at (215:1.16) {\LARGE B};
    \node at (35:1.8cm) {\LARGE $l$};
    \node at (-0.9,1.55) {\LARGE $L$};
    \node at (0,-2.5) {\huge $\beta \propto \sin \left ( \frac{\pi(l-x)}{L}\right ) \sin (\frac{\pi x}{L})$};
    \node at (-2.3,1.5) {\LARGE (a)};

    \shade[shading  = CFT_sin] (7,-0.25) rectangle node[black] {\LARGE \textsc{A}} (10,0.25);
    \node at (7,-2.5) {\huge $\beta \propto \sin (\frac{\pi x}{L})$};
    \draw (4,-0.25) rectangle node[black] {\LARGE \textsc{B}} (7,0.25);
    \draw (7,--0.25) rectangle (10,-0.25);
    \draw [decoration={brace}, decorate, thick] (7.05, 0.325) -- node[midway, above, yshift = 0.05cm] {\LARGE$\sfrac{L}{2}$} (10, 0.325);
    \draw [decoration={brace}, decorate, thick] (4, 0.325) -- node[midway, above, yshift = 0.05cm] {\LARGE$\sfrac{L}{2}$} (6.95, 0.325);
    \node at (4,1.5) {\LARGE(b)};
  \end{tikzpicture}
    }
    \caption{The BW theorem extended to one dimensional systems with conformal symmetry. Schematic representation of the entanglement temperature for (a) a subsystem of length $l$ embedded in a system of length $L$ with periodic 
    boundary conditions and (b) a subsystem of length $\sfrac{L}{2}$ embedded at a boundary of an open system of length $L$.
    The colors indicate a high (orange) and low (blue) entanglement temperature.}
    \label{fig:CFT_extensions}
\end{figure}
In case of a finite subsystem of length $l$ in a ring of circumference $L$, the EH is given by 
\begin{equation}
    \hat{H}_\text{A}^\text{CFT1} = 2L \int_0^l \diff x \, \frac{\sin \left ( \frac{\pi(l-x)}{L}\right ) \sin \left (\frac{\pi x}{L} \right )}{\sin \left (\frac{\pi l}{L} \right )} \hat{\mathcal{H}}(x) + c'. \label{eqn:CFT1}
\end{equation}
Since the system obeys periodic boundary conditions (PBC), there are two entanglement cuts, where the inverse temperature increases approximately linearly for small distances from 
the entanglement cut in agreement with the BW theorem (Eq. \eqref{eqn:BW}).
For a finite partition of length $\sfrac{L}{2}$ at the edge of a finite open system of length $L$,
the EH reads 
\begin{equation}
    \hat{H}_\text{A}^\text{CFT2} = 2L \int_0^{\sfrac{L}{2}} \diff x \, \sin \left ( \frac{\pi x}{L} \right ) \hat{\mathcal{H}}(x) + c', \label{eqn:CFT2}
\end{equation}
again with a linear rise of the inverse entanglement temperature near the entanglement cut (see Fig. \ref{fig:CFT_extensions}).
For a finite subsystem of length $l$ in an infinite composite system, the 
EH is given by 
\begin{equation}
    \hat{H}_\text{A}^\text{CFT3} = 2 \pi \int_0^{l} \diff x \, x \left ( \frac{l-x}{l} \right )  \hat{\mathcal{H}}(x) + c'. \label{eqn:CFT3}
\end{equation} 
With these modified prefactors, we can therefore also describe  conformal extensions, e.g., periodic boundary conditions, beyond the original BW theorem.
\subsection{Quantum classical algorithm}
\label{sec:originalalgorithm}
The main goal of the algorithm, first presented in Ref.~\cite{PhysRevLett.127.170501}, is to learn the EH via a hybrid quantum-classical feedback loop (QCFL) utilizing the variational ansatz 
$\hat{H}_\text{A}^\text{Var}(\bm{g}) = \sum_i g_i \hat{h}_i$, which acts as a generator for the time evolution operator 
\begin{equation}
    \hat{U}_\text{A}(\bm{g},t) = \e^{-\mathrm{i} \var t },
\end{equation}
acting on subsystem A for some time $t$ s.t.
\begin{equation}
    \hat{\rho}_\text{A} \to \hat{U}_\text{A}(\bm{g},t) \hat{\rho}_\text{A} \hat{U}^{\dagger}_\text{A}(\bm{g},t).
\end{equation}
The parameters $g_i$ are the variational parameters of the algorithm. 
\begin{figure}
    \centering
    \resizebox{\columnwidth}{!}{
    \begin{tikzpicture}[scale=1.2]

\draw[thick, orange]    (0,0.5) -- (1,0.5);
\draw[thick, orange]    (0,1)   -- (1,1);
\draw[thick, orange]    (0,1.75)   -- (1,1.75);
\draw[thick, gray]      (0,2.25) -- (1,2.25);
\draw[thick, gray]      (0,2.75)  -- (1,2.75);
\draw[thick, gray]      (0,3.5)  -- (1,3.5);

\draw[thick, orange]    (1.6,0.5)       -- (2,0.5);
\draw[thick, orange]    (1.6,1)         -- (2,1);
\draw[thick, orange]    (1.6,1.75)      -- (2,1.75);
\draw[left color=gray, draw=none] (1.6,2.235) rectangle (2,2.265);
\draw[left color=gray, draw=none]      (1.6,2.735)   rectangle (2,2.765);
\draw[left color=gray, draw=none]      (1.6,3.485)    rectangle (2,3.515);

\draw[thick, orange]    (4.5,0.5) -- (4.9,0.5);
\draw[thick, orange]    (4.5,1)   -- (4.9,1);
\draw[thick, orange]    (4.5,1.75)   -- (4.9,1.75);

\draw[thick, rounded corners=1mm] (1,0.25) rectangle (1.6,3.75);
\node[rotate=90] at (1.3, 2.05) {\Large prepare $\ket{\text{GS}}$};

\node[orange] at (0.15, 1.2) {\Large A};
\node[gray] at (0.15, 2.95) {\Large B};

\node[rotate=90, orange] at (0.5, 1 + 0.75/2) {...};
\node[rotate=90, gray] at (0.5, 2.75+0.75/2) {...};

\draw[thick, rounded corners=1mm] (2,0.25) rectangle (4.5,2);
\node at (3.25, 1.15) {\Large $\e^{-\mathrm{i}\var t_n}$};

\draw[orange, thick] (4.9,0.3) rectangle (5.3,0.7);
\draw[orange, thick] (4.9,0.8) rectangle (5.3,1.2);
\draw[orange, thick] (4.9,1.55) rectangle (5.3,1.95);

\draw[orange, thick] (5.25, 0.35) arc[start angle=0, end angle=180, radius=0.15];
\draw[orange, thick] (5.25, 0.85) arc[start angle=0, end angle=180, radius=0.15];
\draw[orange, thick] (5.25, 1.6) arc[start angle=0, end angle=180, radius=0.15];
\draw[arrows={-Triangle[scale=0.5]}, orange, thick] (5.1, 0.35) -- (5.25,0.64);
\draw[arrows={-Triangle[scale=0.5]}, orange, thick] (5.1, 0.85) -- (5.25,1.14);
\draw[arrows={-Triangle[scale=0.5]}, orange, thick] (5.1, 1.6) -- (5.25,1.89);

\draw[arrows={-Stealth[round]}, blue, line width=1mm] (5.5, 0.3+1.65/2) -- (7.,0.3+1.65/2);
\node at (6.2,0.3+1.65/2 + 0.4) {\Large data};

\draw[thick, rounded corners=1mm] (7.1,0.5) rectangle (8.8,1.75);
\node at (7.95, 0.5+1.75/2-0.5/2+0.2) {\Large optimize};
\node at (7.95, 0.5+1.75/2-0.5/2-0.2) {\Large $\mathcal{C}(\bm{g})$};

\draw[arrows={-Stealth[round]}, blue, line width=1mm]
                          (7.95,1.85) -- (7.95,3) -- (3.25,3) -- (3.25, 2.1);

\node at (5.6, 3.3) {\Large new parameters $g$};

\end{tikzpicture}
    }
    \caption{Quantum classical feedback loop (QCFL). The composite system is initialized 
    with the ground state, $\ket{\text{GS}}$, of the system Hamiltonian.
    The subsystem A is then evolved under the variational ansatz \var and some observables 
    $\langle\obs_j\rangle_{t_n}$ are measured at time instances $\{t_n\}$.
    The cost function is then evaluated with the measurements and the new parameters 
    suggested by the optimizer are used to repeat the procedure.}
    \label{fig:circuit}
\end{figure}
The QCFL, see Fig.~\ref{fig:circuit}, works as follows:
\begin{enumerate}
    \item Prepare an initial state $\hat{\rho}_\text{A} = \Tr_\text{B} \left [\ket{\text{GS}}\bra{\text{GS}} \right ]$ with $\ket{\text{GS}}$ as the ground state of the composite system.
    \item Evolve the subsystem A under the variational ansatz for some time $t_n>0$, leaving the complementary subsystem untouched.
    \item Evaluate the expectation values $\langle \obs_j \rangle_{t_n}$ after each time $t_n$.
    \item Calculate a suitable cost function $\mathcal{C}(\bm{g})$ on a classical computer.
    \item Repeat step 2 to 4 for different variational parameters and minimize $\mathcal{C}(\bm{g})$.
\end{enumerate}
The expectation value after the subsystem A has been evolved under the variational ansatz reads 
\begin{equation}
    \langle\obs_j\rangle_{t_n} = \Tr_\text{A} \left [ \obs_j \hat{U}_\text{A}(\bm{g},t_n) \hat{\rho}_\text{A} \hat{U}^{\dagger}_\text{A}(\bm{g},t_n) \right ],
\end{equation}
where the operators $\obs_j$ are only defined on subsystem A and are restricted to be (quasi-)local.
The optimal parameters \gopt are learned by minimizing the time variation of the observables s.t. 
$\langle\obs_j\rangle_{t_n} = \text{const}$.
A suitable cost function to be minimized is given as
\begin{equation}
    \mathcal{C}(\bm{g}) = \sum_{j=1}^{N_\text{O}}\sum_{n=1}^{N_T} \left (\langle\obs_j\rangle_{t_n} - \langle\obs_j\rangle_0 \right )^2 \label{eqn:cost}
\end{equation}
with $N_\text{O}$ as the number of observables and $N_T$ as the number of times the subsystem A is evolved and each observable is measured.  
For sufficiently many observation times $t_n$ and observables $\obs_j$, a cost function value of zero implies 
\begin{equation}
    \comm{\varopt}{\hat{H}_\text{A}} = 0, \label{eqn:comm}
\end{equation}
where $\hat{H}_\text{A}$ is the exact EH and \gopt are the optimal variational parameters.
Equivalently, a cost function value of zero implies $\comm{\varopt}{\hat{\rho}_\text{A}} = 0$, too, since the exact reduced density matrix $\hat{\rho}_\text{A} = \exp (- \hat{H}_\text{A})$ 
is given by a power series in $\hat{H}_\text{A}$. 
This results in a thermalized subsystem A and the observables are constant in time.
The precise choice of observables is not crucial, since an operator is expected to evolve into a complex operator under the dynamics as long as 
$\comm{\var}{\obs_j} \neq 0$.
Since the aforementioned commutator is still fulfilled if a solution \gopt is scaled by a factor $\gamma$, the scale factor remains undetermined by the algorithm 
as well as the normalization constant $c'$ (see Eq. \eqref{eqn:discrBW}).
To compare the ES of the variational solution and the exact ES, the universal ratios
\begin{equation}
    \kappa_\alpha = \frac{\xi_\alpha - \xi_{\alpha_0}}{\xi_{\alpha_1} - \xi_{\alpha_0}} \label{eqn:universalratios}
\end{equation}
are defined s.t. the undetermined scaling factor $\gamma$ and the normalization constant $c'$ are eliminated by division and subtraction, respectively.
Our numerical simulations use ground states from exact diagonalization and exact time evolution. In quantum simulators based on trapped ions, the entanglement Hamiltonian was determined for a long-range transverse-field Ising model \cite{Joshi2023} and the ground state was prepared variationally. On digital quantum devices, the required time evolution can, in principle, be realized using a Trotter–Suzuki decomposition. Our convergence analysis in Appendix~\ref{sec:conv_analysis} shows that a modest maximum evolution time is sufficient for the models considered. Moreover, as demonstrated in Section \ref{sec:noise_ana}, advanced quadrature schemes require only a small number of time points. Taken together, these observations indicate that the protocol can be implemented with a shallow Trotter–Suzuki decomposition.
\subsection{Improvement of the cost function} 
\label{sec:improvedcost}
Throughout the investigation of the algorithm, some difficulties, and thus, possibilities to improve the algorithm have been noticed.
A major challenge is to determine the observation times $t_n$, such that the optimized parameters are stable to small changes in the $t_n$. This can be easily solved by reinterpreting the cost function.
Recalling the cost function in Eq. \ref{eqn:cost},    
it is hard to compare numerical values of the cost function, since it is not normalized to the number of observables $N_\text{O}$ 
and to the number of observation times $N_T$, which is easily fixed by dividing by these aforementioned quantities.
Since the algorithm is based on monitoring observables, it is, in general, not enough to choose a few arbitrary discrete time points. 
Otherwise, the variational parameters \gopt will not be converged. 
Assuming equidistant time points i.e. a step size $\mathrm{\Delta} t$ for the observation times, 
the cost function can therefore, together with the aforementioned normalization, be rewritten as 
\begin{equation}
    \mathcal{C}(\bm{g}) = \frac{\mathrm{\Delta} t}{T_\text{max} N_\text{O}}\sum_{j=1}^{N_\text{O}}\sum_{n=1}^{N_T} \left (\langle\obs_j\rangle_{n \mathrm{\Delta} t} - \langle\obs_j\rangle_0 \right )^2, \label{eqn:rightpoint}
\end{equation}
defining the maximum observation time $T_\text{max} = N_T\mathrm{\Delta} t$.
To obtain a cost function, observing not at discrete time points but at all times, the discrete sum is replaced by an integral,
\begin{align}
    \mathcal{C}(\bm{g} )& \overset{\overset{\lim}{\mathrm{\Delta} t  \to 0}}{=} 
    \frac{1}{T_\text{max}} \int\limits_0^{T_\text{max}} \underbrace{ \frac{1}{N_\text{O}}\sum_{j=1}^{N_\text{O}} \left (\langle\obs_j\rangle_{t} - \langle\obs_j\rangle_0 \right )^2 }_{\coloneqq c(\bm{g},t)} \diff t  \nonumber \\
    &= \frac{1}{T_\text{max}} \int\limits_0^{T_\text{max}} c(\bm{g},t) \diff t \, , \label{eqn:costcont}
\end{align}
which allows numerical integration techniques to be applied.
Note that there remains one degree of freedom to choose properly, namely the maximum observation time $T_\text{max}$.
The influence of \TMAX will be thoroughly discussed in Appendices \ref{sec:tmax} and \ref{sec:conv_bwv}.
\subsection{Variational ansatz schemes}
We mainly distinguish two types of  variational Ansätze in this work.
The first ansatz is the BW-like ansatz, denoted as \BW, which is used in Reference \cite{PhysRevLett.127.170501}.
As the name suggests, it follows the BW theorem, i.e., it is a local deformation of the system Hamiltonian.
The second ansatz, \BWV, is used to show a violation of the BW theorem in lattice models.
In general, both Ansätze are given by a linear combination 
\begin{equation}
    \var = \sum_i g_i \hat{h}_i, \label{eqn:varansatz}
\end{equation}
where $g_i$ is a variational parameter and \hi is a quasi-local few-body operator, which will be referred to as 
a block.

\subsubsection{BW-like ansatz}
The BW theorem predicts that the EH is a spatially deformed version of the system Hamiltonian on a subsystem.
That is, each lattice site $i$ is assigned a block \hi together with one variational parameter $g_i$,
as illustrated in Fig. \ref{fig:lattice}. 
\begin{figure}
    \centering
     \resizebox{\columnwidth}{!}{
    \begin{tikzpicture}
    
    \begin{scope}[scale=3]
    \draw[thick, gray]    (0.1,0)       -- (0.2,0);
    \draw[thick, gray]    (0.55,0)       -- (0.65,0);
    \draw[thick, gray]    (1.35,0)       -- (1.65,0);
    \draw[thick, gray]    (0.85,0)       -- (1.15,0);

    \draw[inner color=green, draw = none] (2.35,-0.05) rectangle (2.65,0.05);
    \draw[inner color=green, draw = none] (2.6,-0.05) rectangle (2.9,0.05);
    \draw[inner color=green, draw = none] (1.85,-0.05) rectangle (2.15,0.05);

    \draw[fill=white, draw = none] (2.45,-0.1) rectangle (2.8,0.1);

    \shadedraw[ball color = gray, draw = none] (0, 0) circle (0.1);
    \shadedraw[ball color = gray, draw = none] (0.75, 0) circle (0.1);
    \shadedraw[ball color = gray, draw = none] (1.25, 0) circle (0.1);
    \shadedraw[ball color = orange, draw = none] (1.75, 0) circle (0.1);
    \shadedraw[ball color = orange, draw = none] (2.25, 0) circle (0.1);
    \shadedraw[ball color = orange, draw = none] (3, 0) circle (0.1);

    \node[gray, scale = 3]     at (0.75/2, 0) {...};
    \node[orange, scale = 3]   at (2.25+0.75/2, 0) {...};
\end{scope}

\begin{scope}[scale=1]

\node[gray]     at (3*0.6125, -0.2*3) {\Large B};
\node[orange]   at (3*2.5, -0.2*3) {\Large A};

\node at(3*1.75, 3*0.2) {\Large $(\hat{h}_1, g_1)$};
\node at(3*2.25, 3*0.2) {\Large $(\hat{h}_2, g_2)$};
\node at(3*3, 3*0.2)    {\Large $(\hat{h}_{\NA}, g_{\NA})$};

\end{scope}

\end{tikzpicture}
    }
    \caption{Schematic illustration of the variational ansatz $\BW = \sum_i g_i \hi$.
    Each lattice site in the subsystem A is assigned a few-body quasi-local operator \hi together 
    with a variational parameter $g_i$.
    Only interactions within the subsystem A are taken into account, as suggested by the green highlighting.}
    \label{fig:lattice}
\end{figure}
That means that the index $i$ in Eq. \eqref{eqn:varansatz} coincides with the $i$-th lattice site in the subsystem A.
The blocks \hi are not local and act on more than one qubit. 
It is important to note that all interactions are restricted to be within subsystem A as well.
\subsubsection{BW-violating ansatz}
The BW-violating ansatz is not given by a spatially deformed Hamiltonian.
Thus, each lattice site is assigned multiple blocks \hi and multiple variational parameters $g_i$
and the index $i$ in Eq. \eqref{eqn:varansatz} does not coincide with the lattice site $i$.
From now on, the dependence of the variational Ansätze on the variational parameters $\bm{g}$ will be omitted.
\subsection{Transverse field Ising model}
\label{sec:TFIM}
The Hamiltonian of the TFIM with $N$ sites, open boundary conditions (OBC), nearest neighbour coupling strength $J$ and transverse field strength $\Gamma$ reads\cite{PhysRevB.98.134403}
\begin{equation}
    \hat{H} = - J \sum_{i=1}^{N-1} Z_i Z_{i+1}- \Gamma \sum_{i=1}^N X_i,
\end{equation}
where $X_i,Y_i,Z_i$ denotes the Pauli matrices on site $i$.
The first term favors a ferromagnetic state for $J>0$ and an antiferromagnetic state for $J<0$ while 
the transverse field introduces fluctuations s.t. an orientation along the $x$-axis is favored by the 
transverse term.
It possesses a $\mathbb{Z}_2$ symmetry, where the Hamiltonian is invariant 
under flipping all spins, i.e. 
\begin{equation}
    Z_i \to - Z_i .
\end{equation}
In the limit $J \gg \Gamma$, the ground state is two-fold degenerate
and the system is fully polarized with all spins pointing either up or down
\begin{equation}
    \ket{\text{GS}} = \underset{i=1}{\overset{N}{\otimes}} \ket{\uparrow} \quad \text{or}  \quad \ket{\text{GS}} = \underset{i=1}{\overset{N}{\otimes}} \ket{\downarrow}, 
\end{equation}
breaking the $\mathbb{Z}_2$ symmetry spontaneously,
whereas all spins are completely aligned in the $x$-direction in the limit $\Gamma \gg J$
\begin{equation}
    \ket{\text{GS}} = \underset{i=1}{\overset{N}{\otimes}} \underbrace{\frac{1}{\sqrt{2}} ( \ket{\uparrow} + \ket{\downarrow})}_{=\ket{\rightarrow}},   
\end{equation}
exhibiting a paramagnetic behaviour.
The TFIM has a quantum critical point at $\sfrac{J}{\Gamma} = 1$, separating the ordered ferromagnetic and the disordered paramagnetic phase \cite{benedikt}. 
From now on, $J = 1$ holds.
In the case of the ansatz \BW, 
one block of site $i$ is given by 
\begin{equation}
    \hat{h}_i = - \frac{1}{2} \sum_{j \in \langle j,i \rangle \cap \text{A}} Z_j Z_i - \Gamma X_i, 
\end{equation}
where $\langle j,i \rangle \cap \text{A}$ denotes nearest neighbour coupling only if $i$ and $j$ are in the subsystem A.
The BW-violating ansatz for the TFIM is given by 
\begin{equation}
    \BWV = - \sum_{i=1}^{\NA -1} J_{i,i+1}Z_i Z_{i+1} - \Gamma \sum_{i=1}^{\NA} \Gamma_i X_i.
\end{equation}
with $J_{i,i+1}$ and $\Gamma_i$ as variational parameters.
An analytical expression for the EH in the thermodynamic limit in case of a bipartition is given in Reference \cite{https://doi.org/10.1002/andp.202200064}. 
\subsection{XXZ model}
\label{sec:XXZ}
The Hamiltonian of the XXZ model with $N$ lattice sites and OBC is defined as
\begin{equation}
    \hat{H} = \sum_{i=1}^{N-1} \left (  X_i X_{i+1} + Y_i Y_{i+1} + \increment Z_i Z_{i+1} \right ) ,
\end{equation}
where $\increment$ is the anisotropy. 
For $\increment = 1$, the isotropic case, the Heisenberg model is recovered.
The XXZ model is ferromagnetic for $\increment < -1$, quantum critical for $ -1 < \increment  \leq  1 $
, exhibiting a Luttinger liquid phase, and antiferromagnetic for $\increment > 1$ \cite{PhysRevB.98.134403}.
The phase transition at $\increment = -1$ is of first order, s.t. the ferromagnetic state is exact for $\increment < -1$,
while the phase transition at $\increment = 1$ is of second order \cite{PhysRevA.104.012418}.
Again in the ferromagnetic phase, the $\mathbb{Z}_2$ symmetry is spontaneously broken \cite{PhysRevB.98.134403}.  
One block for the ansatz \BW reads
\begin{equation}
    \hat{h}_i = \frac{1}{2} \sum_{j \in \langle j,i \rangle \cap \text{A}} ( X_i X_j + Y_i Y_j + \increment Z_i Z_j), \label{eqn:hiXXZ}
\end{equation}
while the variational ansatz \BWV for the XXZ model reads
\begin{align}
    \BWV = \sum_{i=1}^{\NA-1}  ( & J_{i,i+1}^\text{XX}  (  X_i X_{i+1} + Y_i Y_{i+1}  ) \nonumber  \\ +  & J_{i,i+1}^\text{Z} \increment Z_{i}Z_{i+1} ).
\end{align}
\section{Noise analysis and trainability} \label{sec:noise_ana}
In exact numerical simulations, interpreting the cost function as an integral leads to a substantial improvement in convergence (see Appendix \ref{sec:integral_conv}). For devices in the noisy intermediate-scale quantum (NISQ) regime \cite{Preskill2018quantumcomputingin}, the relevant question is whether this improvement persists in the presence of noise. 

Our numerical simulations are implemented in Julia; technical details are provided in Appendix~\ref{sec:methods}. We provide a Julia package, with full code documentation in \cite{QuantVarEntHam} and the scripts for the plots in \cite{kind_2026_18633065}.

\subsection{Noisy integration}
We model noise at the level of expectation values of the observables entering the cost function. Specifically, for an observable $\obs_j$ evaluated at time $t$, we assume fluctuations with variance 
\begin{align}
\label{eqn:devicenoise}
 \sigma^2 = \text{Var}(\obs_j)_t \sigma^2_\text{device},
\end{align}
where
$\text{Var}(\obs_j)_t= \langle (\obs_j)^2 \rangle_t - \langle \obs_j \rangle_t^2$. 
Here $\sigma^2_\text{device}$ parametrizes the variance of Gaussian noise with zero mean that accounts for the cumulative impact of device noise, including readout error and gate errors. For pure sampling noise we set $\sigma^2_\text{device} = 1/N_\text{M}$, where $N_\text{M}$ is the number of measurements \cite{Kokail2021}. We perform multiple simulations and scan across $\sigma_\text{device}$.

\begin{figure}[t]
    \centering
    \includegraphics[width = \columnwidth]{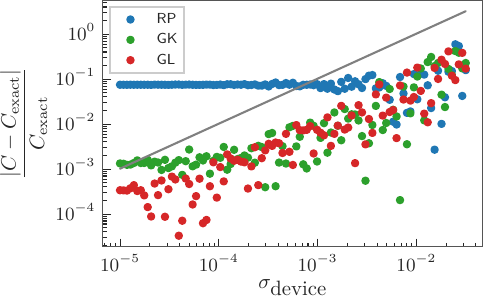}
    \caption{Deviation of the cost function value as a function of $\sigma_\text{device}$ with five time points for the cost function evaluation.
    GK uses the order $n=2$.
    The gray line is given by Eq. \ref{eqn:devicenoise}, where we use the maximum variance over all observables and time points to obtain a 
    conservative upper bound for the uncertainty. The Tanh-sinh method requires at least $13$ function evaluations and we therefore do not compare it here.
    We use the TFIM with $N=8$, $\NA = 4$, $\Gamma = 1$, open boundary conditions with an unoptimized BW-like ansatz.}
    \label{fig:dev_vs_NM}
\end{figure}

We use the simple right point rule (RP) as in \cite{PhysRevLett.127.170501}, the Tanh-sinh method (TS), the adaptive Gauss-Kronrod (GK) and the Gauss-Legendre (GL) quadratures.
For a brief introduction to the integration methods, we refer to Appendix \ref{sec:integration}. We compare the integrators during optimization in Fig.\ \ref{fig:dev_vs_NM} for the TFIM.

For low to intermediate noise levels, the right-point rule exhibits a clear plateau, saturating at a $\sim 7\%$ error, while the Gauss–Kronrod and Gauss–Legendre quadratures converge to values at the noise floor down to $\sigma_\text{device} = 10^{-5}$, with only five function evaluations. In the very large noise regime, all methods yield comparable results. Thus improved quadrature schemes are uniformly preferable: in the presence of strong noise they perform on par with simpler rules, while for low to intermediate noise they provide substantially improved accuracy at the same evaluation cost.

\begin{figure}[t]
    \centering
    \includegraphics[width = \columnwidth]{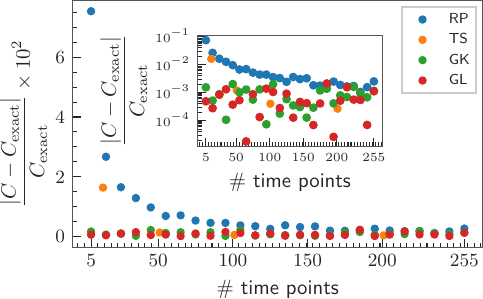}
    \caption{Deviation of the cost function value as a function of the the number of time points for the cost function evaluation. Here, the number of measurements per time point is fixed to $N_\text{M}=\num{e8}$.
    We use the TFIM with $N=8$, $\NA = 4$, $\Gamma = 1$, open boundary conditions with an unoptimized BW-like ansatz.}
    \label{fig:dev_vs_timepoints}
\end{figure}

The importance of a fast converging integration method in the presence of noise is confirmed in Fig. \ref{fig:dev_vs_timepoints}, where we investigate the dependence on the number of function evaluations.
The right point rule reaches the precision of the Gauss-Legendre and Gauss-Kronrod quadrature with five time points at $\sim 250$ time points.
Thus, the right point rule requires $\sim 50$ times more time steps, which in turn requires $\sim 50$ times more measurements to reach the same precision.
\subsection{Analysis of trainability}
The barren plateau problem refers to the concentration of the loss function landscape in variational quantum algorithms, where gradients become exponentially small in system size $N$, rendering optimization ineffective \cite{Larocca2025}. It is characterized by a vanishing mean gradient as well as  an exponentially suppressed variance, indicating that gradients concentrate around zero for almost all parameter choices. We probe this behavior numerically by analyzing the mean and variance of the gradient sampled over random parameters. This statistical characterization provides a direct measure of trainability and also allows us to assess whether increasing expressivity leads to further gradient concentration in the present setting.

Specifically, we compute $680$ gradients sampled uniformly from the interval $[0,N]$ and analyze the mean and variance of their $\ell_\infty$-norm for both the BW-like and the (more expressive) BW-violating ansatz.
We use the TFIM with a varying subsystem chain length $N$, $\NA = \sfrac{N}{2}$, $\Gamma = 1$, OBC, $\TMAX=1$ and $51$ integration steps using Tanh-sinh quadrature.

\begin{figure}[t]
    \centering
    \includegraphics[width = \columnwidth]{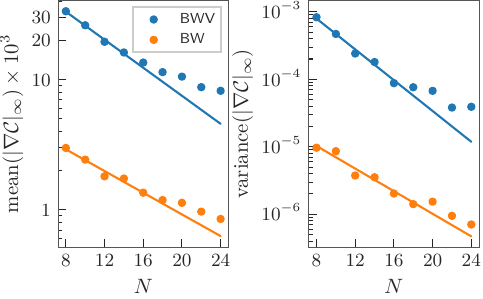}
    \caption{(a) Mean and (b) variance of the $\ell_\infty$-norm of the gradient over $680$ samples for the BW-like (BW) and the BW-violating (BWV) ansatz vs the number of spins in the composite system $N$.
    We use the TFIM with $\NA=N/2$, $\Gamma = 1$ and open boundary conditions.
    Straight lines are exponential fits using system sizes $8, 10, 12,$ and $ 14$.}
    \label{fig:barren}
\end{figure}

In Fig.\ \ref{fig:barren} we show the mean and variance of the gradient in a semi-logarithmic plot. While both quantities decay as a function of system size, this decay is far less severe than the barren-plateau behavior associated with the 2-design properties of random circuits \cite{McClean2018}. For larger qubit numbers the gradient of the BWV ansatz begins to flatten, indicating a sub-exponential decay.

Increased expressivity is often associated with reduced trainability due to loss-function concentration over the enlarged variational manifold \cite{PRXQuantum.3.010313,Arrasmith_2022}. This behavior is not observed in this case. Although the BWV ansatz introduces additional variational parameters, the enlargement of the variational manifold does not lead to a decrease in the gradient. Instead, both mean and variance of the gradient are significantly enhanced, indicating a reduction of loss-function concentration relative to the BW ansatz. This demonstrates that increasing the dimensionality of the variational manifold does not necessarily exacerbate barren-plateau behavior when the added directions are highly relevant for the optimization problem.

\section{Results}
\subsection{Violation of the BW theorem}
\label{sec:BWV_TFIM}

\begin{figure*}[t]
    \centering
    \includegraphics{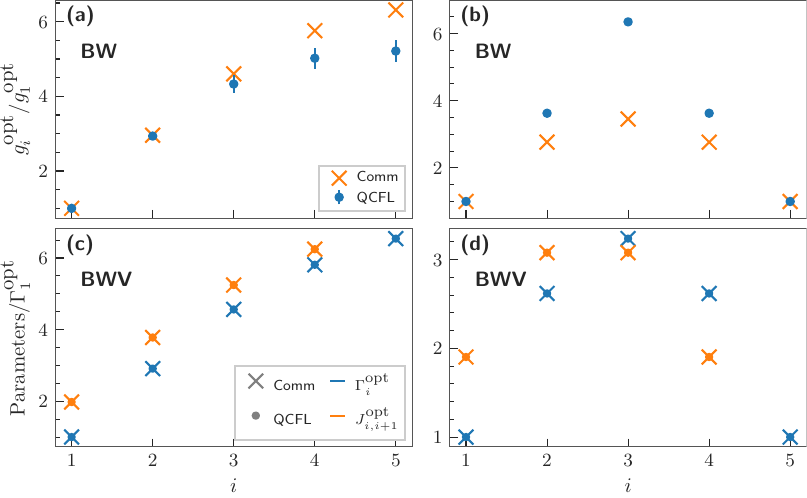}
    \caption{50 runs with different random initial parameters for the TFIM
    with the QCFL and the commutator as a cost function (Comm).
    Outliers from the QCFL were filtered out.
    The mean value from the 50 runs are depicted.
    (a),(b) Optimal parameters for the BW-like ansatz with OBC and 
    PBC, respectively. The error bars for the QCFL are given by the standard deviation over the 50 runs after filtering out. 
    (c),(d) Optimal parameters for the BW-violating ansatz with OBC and 
    PBC, respectively.
    We use the TFIM with $N=10$, $\NA = 5$, and $\Gamma = 1$.}
    \label{fig:BW_vs_BWV_tfim}
\end{figure*}
The convergence properties of different ansatz schemes are analyzed in Appendix \ref{sec:conv_analysis}.
In the following we investigate the variational accuracy of the converged solutions across the different ansatz schemes more systematically by comparing to a cost function that directly measures the commutator between the ansatz Hamiltonian and the exact reduced density matrix.  
We investigate the TFIM with $\Gamma = 1$  using OBC and PBC for the \BW and the \BWV ansatz, with $N=10$, $\NA=5$ and $\TMAX = 1$.
For the same analysis of the XXZ model, we refer to Appendix~\ref{sec:BWV_XXZ}.
We use numerically exact calculations without noise.
We perform a total of 50 minimizations, drawing every initial parameter uniformly from the interval [2, 6].
Outliers are filtered out based on the value of the cost function and the ratios of the optimal parameters.
The optimal parameters from the QCFL are compared to the optimal parameters from the cost function 
\begin{equation}
   \mathcal{C}^\text{Comm}(\bm{g}) = \frac{|| [\hat{\rho}_\text{A}, \varwo (\bm{g}) ] ||_\text{F}}{2||\hat{\rho}_\text{A}||_\text{F}|| \varwo (\bm{g}) ||_\text{F}},
\end{equation}
which measures the commutativity of the exact reduced density matrix and the variational ansatz. 
The notation $||X||_F$ with some $n\times m$ matrix $X$ denotes the Frobenius norm of $X$ defined as\cite{Frob}
\begin{equation}
    ||X||_\text{F} = \left( \sum_{i=1}^{n} \sum_{j=1}^{m} |x_{ij}|^2 \right)^{\frac{1}{2}}.
\end{equation}
Here, $n=m=2^{\NA}$ holds.
No filtering of outliers is done with the results of the aforementioned commutator as a cost function $\mathcal{C}^\text{Comm}(\bm{g})$.
We plot the ratios of the parameters to eliminate the undetermined scale factor in the found solution (see Eq.~\eqref{eqn:universalratios}).

If the ansatz \BW is used in the case of OBC, only the ratio $\sfrac{g_2^\text{opt}}{g_1^\text{opt}}$ obtained with the QCFL shows good agreement with 
the ratio obtained via the commutator (Fig. \ref{fig:BW_vs_BWV_tfim}(a)), whereas the third, fourth, and fifth ratios deviate significantly. 
A linear rise near the entanglement cut can be observed and bending in accordance with the 
second CFT extension (Eq.~\eqref{eqn:CFT2}) becomes apparent.
In the case of PBC, the optimal parameters exhibit large deviations from the parameters obtained via the commutator
and do not follow the first CFT extension (Eq.~\ref{eqn:CFT1}) but rather 
a triangular form, while the parameters yielded with the commutator do show the 
behaviour predicted by the first CFT extension (Fig.~\ref{fig:BW_vs_BWV_tfim}(b)).
In both cases, especially in the case of PBC, the standard deviation is very small, indicating the 
minimizer almost always finds the same solution.
This observation is supported by the cost function value at its minimum agreeing in almost all runs, which 
takes the value of $\copt \approx \num{1.08(2)e-5}$ and 
$\copt \approx \num{3.95e-4}$ for OBC and PBC, respectively.
On the other hand, if the ansatz \BWV is used, the optimal parameters obtained via the QCFL match perfectly with the parameters obtained via 
the commutator for both, OBC and PBC (see Fig. \ref{fig:BW_vs_BWV_tfim}(c),(d)).
In addition to the perfect match, 
the standard deviation of each parameter obtained via the QCFL is in the vicinity of $\num{e-14}$, indicating that 
there is one good solution, while
the standard deviation of the ratios obtained via the commutator are not larger than $\num{6e-9}$, which is why
no error bars are depicted.
The minimum of the cost function is numerically zero for both OBC and PBC, 
underlining the fact that the ansatz \BWV works much better.
This means that the BW theorem does not deliver 
an accurate description for the EH for the TFIM on a finite size chain. 
Additionally, the only corrections to the BW theorem are
that there is not just one parameter per lattice, but two.
Further corrections such as long-range interactions or higher-body interactions are not needed, since 
the cost function is already numerically zero at its minimum.
\subsection{TFIM and XXZ model across the phase diagram}
This section investigates the TFIM and XXZ model across the respective phase diagrams to 
analyze how the algorithm performs when the systems are not critical and whether the algorithm can indicate the critical points or 
specific phases in the presence of measurement uncertainty. 

Note that the statistical error of the cost function is governed by the measurement uncertainty of the observables, modeled as noise $\sigma_{j,t_n}$ in each measured expectation value $\langle \obs_j \rangle_{t_n}$,
\begin{align}
C^{\text{noisy}}
=& \frac{1}{\TMAX} \int_0^{\TMAX} \sum_{j=1}^{N_\text{O}} \frac{1}{N_\text{O}} \left( \Delta\langle \obs_j \rangle_{t} + \sigma_{j,t} \right)^2 \diff t \nonumber \\
=& \frac{1}{\TMAX} \int_0^{\TMAX} \sum_{j=1}^{N_\text{O}} \frac{1}{N_\text{O}} \left ( \Delta\langle \obs_j \rangle_{t}^2 \right. \nonumber \\
 & \left. +2 \Delta\langle \obs_j \rangle_{t}\sigma_{j,t} +\sigma_{j,t}^2 \vphantom{\Delta\langle \obs_j \rangle_{t}} \right ) \diff t, 
\end{align}
where \(\Delta\langle \obs_j \rangle_{t} \equiv \langle \obs_j \rangle_{t} - \langle \obs_j \rangle_{0}\).
We use this scaling to realize a controlled zero-noise extrapolation of the cost function. For each point in the phase diagram, we extrapolate the measured cost to the zero-noise limit using data obtained at different $N_\mathrm{M}$. Details of the extrapolation procedure are provided in Appendix~\ref{sec:ZNE}.

\begin{figure}[t]
    \centering
    \includegraphics[width =\columnwidth]{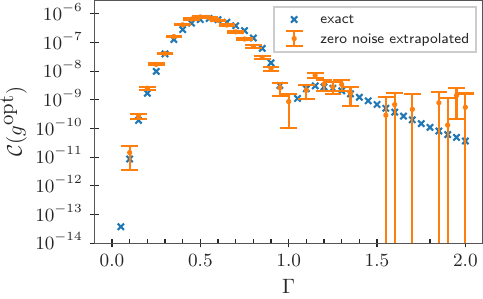}
    \caption{Minimum of the cost function in dependence on the
     transverse field strength $\Gamma$ for the TFIM, when the ansatz 
     \BWV is used. 
     We perform simulations without noise (exact) and noisy simulations with zero-noise extrapolations.
     Values below double precision and extrapolated values below zero are discarded.
     The error bars stem from the uncertainty of the fit parameters.
     We use the TFIM with $N=8$, $\NA = 4$ and open boundary conditions.}
    \label{fig:PDcostTFIM}
\end{figure}

The noisy simulations use the Gauss-Legendre quadrature with five time points. We compare with exact numerics using the fully converged solution with the Tanh-sinh method.
The ansatz \BWV is used without any other corrections for both models and the maximum integration time is set to $\TMAX = 2$.\\
Figure \ref{fig:PDcostTFIM} shows the minimum of the cost function for varying the transverse field strength $\Gamma$ in the TFIM.
At $\Gamma = 0$ the minimum of the cost function is zero, consistent with the fact that the ground state is unentangled.
A second notable point appears at $\Gamma = 1$, the quantum critical point of the model, where the cost function is zero as well. For intermediate values of $\Gamma$ the cost function remains low but takes finite values.
The zero-noise extrapolated values are in close agreement with the exact values. We use a reasonable number of measurements ($\sim10^{9}$), which is sufficient to extract small values of the cost function. This number of measurements is within reach of superconducting quantum processors, see \cite{PhysRevD.109.114510,abanin2025constructiveinterferenceedgequantum}. \\

\begin{figure}[t]
    \centering
    \includegraphics[width =\columnwidth]{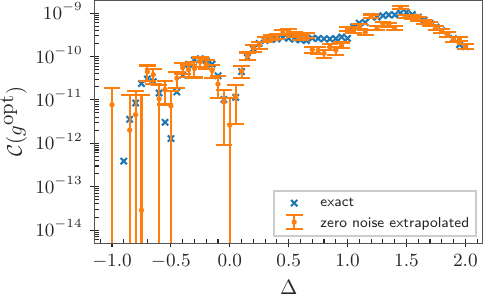}
    \caption{Minimum of the cost function in dependence on the
    anisotropy $\increment$ for the XXZ model, when the ansatz 
     \BWV is used. 
     We perform simulations without noise (exact) and noisy simulations with zero-noise extrapolations.
     Values below double precision and extrapolated values below zero are discarded.
     The error bars stem from the uncertainty of the fit parameter.
     We use the XXZ model with $N=8$, $\NA = 4$ and open boundary conditions.}
    \label{fig:PDcostXXZ}
\end{figure}

\noindent
In the case of the XXZ model, see Fig. \ref{fig:PDcostXXZ}, the first order phase transition at $\increment = -1$ is clearly visible. Here, due to the smaller cost function, we require two orders of magnitude more measurements in the zero-noise extrapolation ($\sim10^{11}$). For $\increment < -1$ the ground state of the XXZ model is a simple product state with all spins pointing in the same direction, and thus, no entanglement is present in the composite system. 
For the system size shown here the second order phase transition at $\increment = 1$ is not visible, but can be seen as a cusp-like behaviour of the cost function for larger system sizes, see Appendix~\ref{sec:xxz_phase}.
For $\increment = 0$ the system is at its self-dual point. In terms of Jordan-Wigner fermions it is non-interacting, and the EH is captured perfectly by the BWV ansatz.
The cost function values show a cusp-like behavior at $\increment=-0.5$, although there is no phase transition. This point is known to be algebraically special in the integrable structure of the XXZ chain.

To conclude, the algorithm delivers indications for the quantum phase transitions and classical states.
Additionally, the accuracy of the BW-violating ansatz varies across the phase diagram.
\subsection{Violation of the BW theorem in the XXZ model in the thermodynamic limit}
\begin{figure}
    \centering
    \includegraphics[width = \columnwidth]{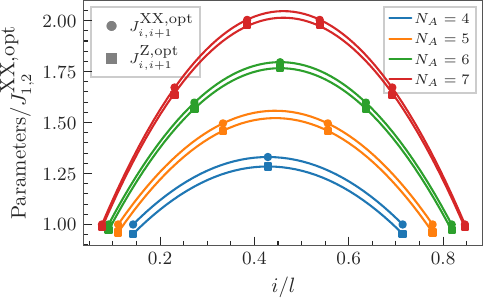}
    \caption{Optimal parameters $J_{i,i+1}^\text{XX,opt}$ and $J_{i,i+1}^\text{Z,opt}$ normalized to $J_{1,2}^\text{XX,opt}$ 
    extrapolated into the TDL vs. the lattice site $i$ in units of the subsystem chain length $l$ for each number of sites \NA in the subsystem A.
    We use $\increment = -0.5$.}
    \label{fig:exXXZ}
\end{figure}

In this section we discuss the reliability of extrapolating the obtained results into the thermodynamic limit (TDL). 
We focus on the XXZ model at $\increment = -0.5$, where violations of the BW theorem have been observed in Ref.~\cite{Nienhuis_2009}.
For simplicity we neglect long-range corrections. The subsystem chain length ranges from $\NA = 4$ to $\NA = 7$,
and $\TMAX = 5$ is used.
To measure the deviation from the BW theorem, the quantity 
\begin{equation}
    \theta_i = \frac{J_{i,i+1}^\text{XX,opt}}{J_{i,i+1}^\text{Z,opt}} - 1 \label{eqn:discr}
\end{equation}
is defined, which will be referred to as the discrepancy.
The procedure to obtain the parameters in the TDL and 
an example plot to show how the fit is done is given 
in Appendix \ref{sec:extr}.
The extrapolated parameters normalized to $J_{1,2}^\text{XX,opt}$ are given in Fig. \ref{fig:exXXZ}, where
the solid lines are there to guide the eye and take the quadratic form $\propto i \frac{\NA - i}{\NA}$, as suggested by the conformal extension 
$\hat{H_\text{A}^\text{CFT3}}$ (see Eq.~\ref{eqn:CFT3}),
although the CFT extensions apply only to the BW theorem.
No error bars are given, since the propagated estimated uncertainties of the parameters of the fit (see Eq.~\eqref{eqn:fitodd}) 
are not larger than than $\num{2e-3}$. 
The gap between the $x$- and $y$-couplings, and the $z$-couplings can be seen.
\begin{table}
    \centering
    \caption{The mean value of the discrepancies as defined in Eq.~\eqref{eqn:discr} in the TDL for each subsystem size \NA.}
    \label{tab:discrepancies}
    \sisetup{table-format=1.4}
    \begin{tblr}{
        colspec = {S[table-format=1.0] S S},
        row{1} = {guard, mode=math},
        vline{3} = {2}{-}{text=\clap{$\pm$}},
        }
        \toprule
        \NA & \SetCell[c=2]{c} \bar{\theta}_i \\
        \midrule
        4   & 0.0443 & 0.0002   \\
        5   & 0.0326 & 0.0006  \\
        6   & 0.0224 & 0.0004   \\
        7   & 0.0154 & 0.0006   \\
        \bottomrule
    \end{tblr}
\end{table}
Table \ref{tab:discrepancies} lists the mean value of the discrepancies, as defined in Eq.~\eqref{eqn:discr}, over the lattice sites $i$ for 
all subsystem lattice sizes \NA in the TDL.
The errors given in Table \ref{tab:discrepancies} are the propagated errors, stemming from the estimated uncertainties of the fit.\\
We conclude that a clear discrepancy can be observed in the TDL.
However, instead of the discrepancies of $\approx 0.1$ from Reference \cite{Nienhuis_2009},
the discrepancies found with the algorithm of this work are approximately two times lower.
\subsection{Comparison of the Entanglement spectra}
\label{sec:entspectra}
\begin{figure}[t]
    \centering
    \includegraphics[width = \columnwidth]{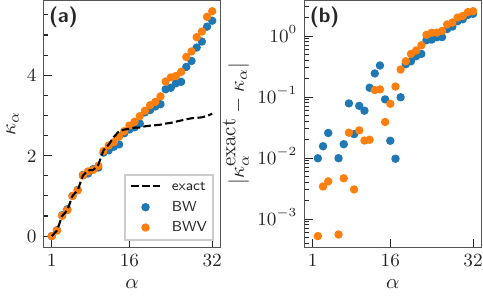}
    \caption{(a) Universal ratios
    and (b) deviations from the exact universal ratios for the BW-like ansatz \BW (BW) and 
    BW-violating ansatz \BWV (BWV) in the TFIM.
    We use $N=10$, $\NA = 5$, $\Gamma = 1$ and open boundary conditions.}
    \label{fig:spectra_TFIM}
\end{figure}

This section compares the universal ratios (see Eq.~\eqref{eqn:universalratios}) of the 
variational solutions of the previous sections to the exact ratios.
We choose $\alpha_0 = 1$ and $\alpha_1 = 5$.
The exact ES is computed through exact diagonalization of the exact EH, given by $\hat{H}_\text{A} = - \ln \left (  \hat{\rho}_\text{A} \right ) $.\\
In the case of the TFIM, the low-lying universal ratios match the exact universal ratios better if 
the BW-violating ansatz is used, as expected (see Fig. \ref{fig:spectra_TFIM} (a),(b)).
In contrast, in the higher part of the spectrum, the universal ratios obtained via variation show significant deviations from the 
exact universal ratios (Fig. \ref{fig:spectra_TFIM}(b)).
One source for this discrepancy is numerical precision: the reduced density matrix contains (eigen-)values, which are so small s.t. they cannot be accurately captured with double precision.
Since the ES (and thus the universal ratios) are obtained by taking the logarithm of the 
reduced density matrix, the lowest eigenvalues of the reduced density matrix are mapped to the highest universal ratios. 
Thus, to compare the universal ratios, we focus on the first few universal ratios.
The first ten universal ratios exhibit a mean absolute deviation from the exact universal ratios 
of $\overline{\increment \kappa^\text{BW}_{\alpha}} = 0.0253$ and  
$\overline{\increment \kappa^\text{BWV}_{\alpha}} = 0.0071$ for the ansatz $\BW$ and $\BWV$, respectively.
That is, the low-lying spectrum (here, the first ten universal ratios) is reconstructed more than three times more accurately on average
if the BW-violating ansatz is used.\\
The universal ratios for the XXZ model are shown in Fig. \ref{fig:spectra_XXZ}(a).
Here we include the BW-violating ansatz \BWV with long range corrections (see Appendix~\ref{sec:correctionsXXZ}).
As can be seen in Fig. \ref{fig:spectra_XXZ}(b), the low-lying spectrum is not significantly better reconstructed if the ansatz 
\BWV or the ansatz \BWV with all long-range corrections is used.    
\begin{figure}
    \centering
    \includegraphics[width =\columnwidth]{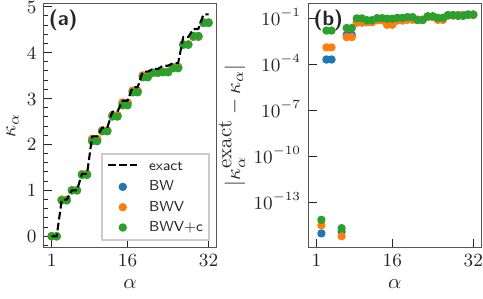}
    \caption{(a) Universal ratios
    and (b) deviations from the exact universal ratios for the BW-like ansatz \BW (BW), 
    BW-violating ansatz \BWV (BWV) and the BW-violating ansatz with all long-range corrections (BWV+c) in the XXZ model.
    We use $N=10$, $\NA = 5$, $\Delta = -0.5$ and open boundary conditions.}
    \label{fig:spectra_XXZ}
\end{figure}
\begin{table*}[t]
    \centering
    \caption{Measures to compare the accuracy of the variational Ansätze, 
    BW-like ansatz (BW), BW-violating ansatz (BWV) and the BW-violating ansatz with all long-range corrections (BWV+c) for the XXZ model.
    We use $N=10$, $\NA = 5$, $\increment=-0.5$ and open boundary conditions.}
    \label{tab:devexact}
    \sisetup{table-format=1.10}
    \begin{tblr}{
        colspec = {c S[table-format=1.6] S[table-format=1.6] S S},
        row{1} = {guard, mode=math},
        }
        \toprule
        \text{ansatz}  & \overline{\increment \kappa_{\alpha}} & \mathcal{T}(\hat{\rho}_\text{A}, \hat{\rho}_\text{A}^\text{Var}) &  \mathcal{F}(\hat{\rho}_\text{A}, \hat{\rho}_\text{A}^\text{Var}) 
        & \copt \\
        \midrule
        \text{BW}       &   0.01309 & 0.00751 & \num{4.17379e-4} & \num{3.55022e-7}\\
        \text{BWV}      &   0.01256 & 0.01983 & \num{9.32176e-5} & \num{5.76821e-10}\\
        \text{BWV+c} &   0.02887 & 0.03030 & \num{1.37317e-6} & \num{1.94540e-15}\\
        \bottomrule
    \end{tblr}
\end{table*}
To understand this behaviour, we investigate two additional measures, which involve the variational reduced density matrix on subsystem A 
\begin{equation}
    \hat{\rho}_\text{A}^\text{Var} = \frac{1}{\Tr \left[ \e^{-\varopt} \right]} \e^{-\varopt}.
\end{equation}
The first measure is the trace distance
\begin{equation}
    \mathcal{T}(\hat{\rho}_\text{A}, \hat{\rho}_\text{A}^\text{Var}) = 
    \frac{1}{2} \Tr \left [ \sqrt{\left( \hat{\rho}_\text{A} - \hat{\rho}_\text{A}^\text{Var} \right)^2} \, \right  ],
\end{equation}
which measures how close two quantum states are 
and ranges from 0 (identical states) to 1 (maximally distant states) \cite{Nielsen_Chuang_2010}.
The second measure utilizes the commutator of $\hat{\rho}_\text{A}$ and 
$\hat{\rho}_\text{A}^\text{Var}$
\begin{equation}
    \mathcal{F}(\hat{\rho}_\text{A}, \hat{\rho}_\text{A}^\text{Var}) = \frac{|| [\hat{\rho}_\text{A}, \hat{\rho}_\text{A}^\text{Var} ] ||_\text{F}}{2||\hat{\rho}_\text{A}||_\text{F}||\hat{\rho}_\text{A}^\text{Var}||_\text{F}}, \label{eqn:commutatormin}
\end{equation}
which, ranges from 0 (completely commuting) to 1 (maximally non-commutative) \cite{4024ce7c-aa06-372d-8878-08ba4cb37bd0}.
This measure is included, since the cost function is based on the commutativity of the variational ansatz and the 
exact reduced density matrix.
Table \ref{tab:devexact} lists the mean of the absolute deviations of the first ten universal ratios from the exact universal ratios 
$\overline{\increment \kappa_{\alpha}}$ , the trace distance $\mathcal{T}(\hat{\rho}_\text{A}, \hat{\rho}_\text{A}^\text{Var})$,
the norm of the commutator $\mathcal{F}(\hat{\rho}_\text{A}, \hat{\rho}_\text{A}^\text{Var})$
and the cost function value at its minimum $\copt$.
It can be seen that $\mathcal{F}(\hat{\rho}_\text{A}, \hat{\rho}_\text{A}^\text{Var})$ is lower if the ansatz \BWV is used and 
the lowest if the long-range corrections are included.
This is in agreement with the cost function value, which shows the same trend.
This observation makes sense, since the cost function is based on the commutativity of the exact reduced density matrix and the variational ansatz.
However, the trace distance is the highest for the ansatz \BWV with long-range interactions included and the lowest
for the BW-like ansatz \BW.
Thus, the universal ratios are \emph{not} reconstructed more accurately with the ansatz \BWV with long-range interactions,
since the trace distance is a measure for how close two quantum states are. We conclude that a cost function based on the commutator between the exact and the variational reduced density matrices is not necessarily sufficient to minimize the trace distance between the variational EH and the exact EH. A zero commutator only implies a shared eigenbasis of the operators and not the same eigenvalues. Any function of the exact EH also leads to stationarity in the dynamics of the subsystem. 
While this clearly restricts the applicability of the algorithm, we expect that certain properties remain invariant, even if the variational algorithm captures only the eigenbasis correctly, such as degeneracies.

\subsection{Topological Entanglement Spectrum}
To investigate degeneracies in the ES,  we consider the spin-1 chain model 
\begin{equation}
    \hat{H} = J \sum_i \hat{\Vec{S}}_i \cdot \hat{\Vec{S}}_{i+1}
    + B_x \sum_i \hat{S}_i^x + U_{zz} \sum_i (\hat{S}_i^z)^2,
\end{equation}
where $\hat{S_i^{\alpha}}$ with $\alpha \in \{x,y,z\}$
are the components of the the spin-1 operator $\hat{\Vec{S}}_i$ on
the $i$-the lattice site.
Pollmann \textit{et al.} \cite{PhysRevB.81.064439} showed that this model 
has a topological Haldane phase, where the ES has symmetry protected degeneracies in case of a semi-infinite half-chain.
We consider a composite system of $N=8$ sites with PBC, 
$J=1$, $B_x = 0$ and $U_{zz} = 0$.
The subsystem consists of $\NA = 4$ sites and we use use the variational ansatz 
\begin{widetext}
\begin{equation}
        \varwo = J \sum_{i=1}^{N_\text{A}-1} ( J_{i,i+1}^x S_i^x S_{i+1}^x + 
    J_{i,i+1}^y S_i^y S_{i+1}^y + J_{i,i+1}^z S_{i}^z S_{i+1}^z ) 
    + B_x \sum_{i=1}^{N_\text{A}} B_i^x S_i^x  
     +  U_{zz} \sum_{i=1}^{N_\text{A}} U_i^{zz} (S_i^z)^2,
\end{equation}
\end{widetext}
where $\{J_{i,i+1}^x, J_{i,i+1}^y, J_{i,i+1}^z, B_i^x, U_i^{zz}\}$
are the variational parameters.
\begin{figure}
    \centering
    \includegraphics[width =\columnwidth]{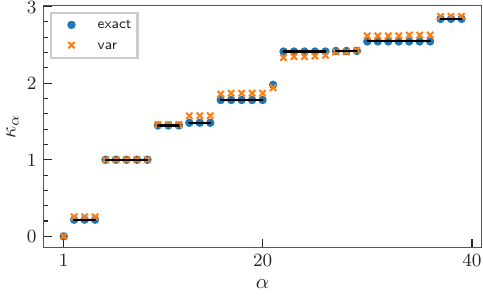}
    \caption{First forty variational (var) and exact universal ratios with 
    $N=8$, PBC, $J=1$, $B_x = 0$, $U_{zz} = 0$ and $\NA = 4$.
    The cost function value at the minimum is $\copt \approx \num{5.29e-05}$.
    The black lines indicate which eigenvalues are degenerate.}
    \label{fig:topo}
\end{figure}
Figure \ref{fig:topo} shows the first forty values of the exact and variational ES.
The cost function converged to 
$\copt \approx \num{5.29e-05}$.
Although the cost function is still finite, the variational algorithm can reproduce the degeneracies in the ES.
We do not find the perfect double degeneracy in the ES, as predicted for the system in the TDL \cite{PhysRevB.81.064439}, due to finite size effects.
We conclude that degeneracies can be obtained with the variational algorithm, even if the minimum of the cost function is finite.
\section{Conclusion and Outlook}
In summary, the discretized Bisognano–Wichmann Ansatz provides a useful starting point for learning lattice entanglement Hamiltonians, but allowing independent couplings per site already improves the variational description markedly, while in anisotropic Heisenberg chains a set of long-range corrections is still necessary. A system size analysis of the gradient demonstrates that the BW-violating ansatz has improved trainability despite a greater expressivity compared to the BW ansatz. This is a counterexample to what is often expected from more expressive circuits. Recasting the cost function as a time integral and evaluating it with iterative quadrature lowers the sampling effort by orders of magnitude in the absence of noise, and still by a factor of $50$ in the presence of noise, a saving that will be indispensable when the protocol is run on noisy quantum hardware.  The cost itself carries physical information, as it shows pronounced minima at phase transitions and remains finite otherwise, so that the cost function itself emerges as a simple experimental probe.  Because the optimization enforces commutativity with the reduced density matrix, the learned operator reproduces the eigenbasis of the exact entanglement Hamiltonian and therefore captures degeneracies and gaps in the spectrum, even though individual eigenvalues are obtained only qualitatively.

An immediate outlook is to implement the iterative integration procedure and the BW violating Ansatz on analogue or digital simulators to demonstrate the improvement resulting from the integral reformulation of the cost function.  Finally, the hybrid loop is not tied to ground-state physics. Applying it to excited, driven and Floquet states, where more expressive Ansatz schemes are necessary, should open the way to entanglement spectroscopy of excited states and non-equilibrium phases on current devices.

\clearpage

\onecolumn

\section*{Acknowledgements}
We thank Peter Zoller, Christian Kokail and Bhuvanesh Sundar for helpful  comments. This work was funded by the Deutsche Forschungsgemeinschaft (DFG, German Research Foundation)-Project No. FA 1884/5-1. Q-Neko project has received funding from the European Union’s Horizon Europe research and innovation programme under Grant Agreement No. 101241875. This work was also performed for Council for Science, Technology and Innovation (CSTI), Cross-ministerial Strategic Innovation Promotion Program (SIP), “Promoting the application of advanced quantum technology platforms to social issues”(Funding agency: QST)

\bibliographystyle{quantum}

\onecolumn

%\bibliography{references.bib}

\clearpage

\appendix

\section{Convergence analysis}
\label{sec:conv_analysis}
In this section we analyze the dependence of the cost function of the BW-like Ansatz on the number of observed time points, the maximum observation time and we discuss the convergence properties of the BW-violating Ansatz. 
Throughout this section, the TFIM with $N=8$, $\NA = 4$, $\Gamma = 1$ and OBC is used.
\subsection{Cost function and convergence of the midpoint rule
for the BW-like Ansatz}
\begin{figure}[h]
    \centering
    \includegraphics{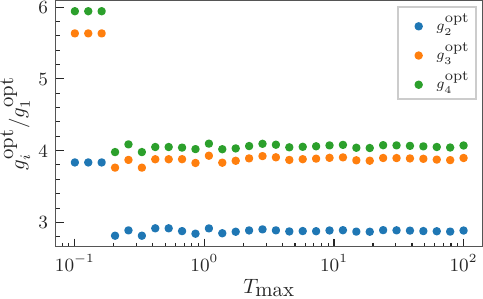}
    \caption{Optimal parameters normalized to $g_1^\text{opt}$ vs. \TMAX for the Ansatz \BWV. For each \TMAX the same initial parameters have been used.
    We use $N=8$, $\NA = 4$, $\Gamma = 1$ and open boundary conditions.}
    \label{fig:paramsvstmax_noconsec}
\end{figure}
\label{sec:integral_conv}
Table \ref{tab:parameters_midpoint} contains the normalized optimal parameters $\gopt/g^\text{opt}_1$ rounded to 
15 decimal places obtained via the midpoint rule with different \dt and the Tanh-sinh quadrature (see appendix \ref{sec:integration}) and the corresponding cost function value \copt.
Here, the BW-like Ansatz \BW is used.
The maximum observation time is $\TMAX = 1$.
All runs are initialized with the same initial parameters.\\
\begin{table*}[h]
    \centering
    \caption{Normalized optimal parameters $\sfrac{\gopt}{g_1^\text{opt}}$ for the Ansatz \BW obtained with the midpoint rule with different \dt and 
    the Tanh-sinh quadrature (101 evaluations).  All runs were initialized with the same initial parameters.
    We use the TFIM with $N=8$, $\NA = 4$, $\Gamma = 1$ and open boundary conditions.}
    \label{tab:parameters_midpoint}
    \sisetup{table-format=1.12}
    \begin{tblr}{
        colspec = {c S S S c},
        row{1} = {guard, mode=math},
        }
        \toprule
        \text{method} & g_2^\text{opt}/g_1^\text{opt} & g_3^\text{opt}/g_1^\text{opt} & g_4^\text{opt}/g_1^\text{opt} & \copt \\
        \midrule
        \dt  = \num{e-1}    & 3.824039596427     & 5.608777483165     &   5.921893959512  & \num{2.5110e-05} \\
        \dt  = \num{e-2}    & 3.835351726582     & 5.632970094804     &   5.941793208331  & \num{2.0946e-05}\\
        \dt  = \num{e-3}    & 3.835421398527     & 5.633117585462     &   5.941909050130  & \num{2.0948e-05}\\
        \dt  = \num{e-4}    & 3.835422093758     & 5.633119057108     &   5.941910205525  & \num{2.0948e-05}\\
        \dt  = \num{e-5}    & 3.835422100710	    & 5.633119071825     &   5.941910217079  & \num{2.0948e-5}\\
        \dt  = \num{e-6}    & 3.835422100782	    & 5.633119071975     &   5.941910217197  & \num{2.0948e-05} \\
        \text{Tanh-sinh}    & 3.835422100780     & 5.633119071973     &   5.941910217195  & \num{2.0948e-05}\\
        \bottomrule
    \end{tblr}
\end{table*} 
\begin{table*}[t]
    \centering
    \caption{Optimal parameters normalized to $\Gamma_1^\text{opt}$ for the Ansatz \BWV and its corresponding minimum of the cost function for 
    different time steps \dt for the midpoint rule.
    All runs were initialized with the same initial parameters.
    We use the TFIM with $N=8$, $\NA = 4$, $\Gamma = 1$ and open boundary conditions.}
    \label{tab:midpoint_BWV_dt}
    \sisetup{table-format=1.12}
    \begin{tblr}{
        colspec = {c S S S},
        row{1} = {guard, mode=math},
        }
        \toprule
        \text{method} & J_{1,2}^\text{opt} / \Gamma_1^\text{opt} & \Gamma_2^\text{opt} / \Gamma_1^\text{opt}& J_{2,3}^\text{opt}/ \Gamma_1^\text{opt}\\
        \midrule
        \dt  = \num{e-1} & 1.965946199368   & 2.864944458809    & 3.666380470827 \\
        \dt  = \num{e-2} & 1.965946199368   & 2.864944458809    & 3.666380470827 \\
        \dt  = \num{e-3} & 1.965946199368   & 2.864944458809    & 3.666380470827 \\
        \dt  = \num{e-4} & 1.965946199368   & 2.864944458809    & 3.666380470827 \\
        \dt  = \num{e-5} & 1.965946199368   & 2.864944458809    & 3.666380470827 \\
        \text{Tanh-sinh} & 1.965946199368   & 2.864944458809    & 3.666380470827 \\
    \end{tblr}
    \begin{tblr}{
        colspec = {S S S c},
        row{1} = {guard, mode=math},
        }
        \toprule
        \Gamma_3^\text{opt}/ \Gamma_1^\text{opt} & J_{3,4}^\text{opt} / \Gamma_1^\text{opt}& \Gamma_4^\text{opt} / \Gamma_1^\text{opt} & \mathcal{C}(\gopt) \\
        \midrule
        4.342962293250 & 4.871649743586   & 5.234439004803 & \num{0}   \\
        4.342962293250 & 4.871649743585   & 5.234439004803 & \num{0}   \\
        4.342962293250 & 4.871649743585   & 5.234439004803 & \num{0}   \\
        4.342962293250 & 4.871649743585   & 5.234439004803 & \num{0}   \\
        4.342962293250 & 4.871649743586   & 5.234439004803 & \num{0}   \\
        4.342962293250 & 4.871649743586   & 5.234439004803 & \num{0}   \\
        \bottomrule
    \end{tblr}
\end{table*}
Here, the fact that the cost function value at the minimum \copt is not numerically zero is crucial.
It can be seen that the optimal parameters obtained with the midpoint rule approach the optimal parameters 
calculated with the Tanh-sinh quadrature asymptotically from below as the time step size \dt decreases, except for $\dt = \num{e-6}$, where
the ratios exceed those obtained with the Tanh-sinh quadrature.
Note that for $\dt = \num{e-2}$ the integrand is evaluated $100$ times but only two decimal places match the results 
of the Tanh-sinh quadrature, while for $\dt = \num{e-6}$ eleven decimal places are in agreement with the Tanh-sinh quadrature but 
the integrand is evaluated at $100000$ time points.
The Tanh-sinh quadrature is more efficient, since it evaluates the integrand only $101$ times at the minimum of the cost function
and gives highly accurate results.
These results demonstrate that, to get accurate results with the BW-like Ansatz, it is not sufficient to observe the observables at a few arbitrary time points, underlining that the cost function should not be treated as a discrete sum over a few time points but rather as an integral over the time domain.
\subsection{Influence of the maximum observation time for the BW-like Ansatz}
\label{sec:tmax}

\begin{table*}[t]
    \centering
    \caption{Optimal parameters normalized to $\Gamma_1^\text{opt}$ for the Ansatz \BWV and its corresponding minimum of the cost function for 
    different \TMAX.
    The Tanh-sinh quadrature was used to evaluate the cost function.
    All runs were initialized with the same initial parameters.
    We use the TFIM with $N=8$, $\NA = 4$, $\Gamma = 1$ and open boundary conditions.}
    \label{tab:midpoint_BWV_Tmax}
    \sisetup{table-format=1.12}
    \begin{tblr}{
        colspec = {c S S S},
        row{1} = {guard, mode=math},
        }
        \toprule
        \TMAX & J_{1,2}^\text{opt}/ \Gamma_1^\text{opt} & \Gamma_2^\text{opt}/ \Gamma_1^\text{opt} & J_{2,3}^\text{opt}/ \Gamma_1^\text{opt}\\
        \midrule
        0.1 & 1.965946199360 & 2.864944458793 & 3.666380470804   \\
        1   & 1.965946199368 & 2.864944458809 & 3.666380470827 \\
        10  & 1.965946199368 & 2.864944458809 & 3.666380470827 \\
        100 & 1.965946199368 & 2.864944458809 & 3.666380470827 \\
    \end{tblr}
    \begin{tblr}{
        colspec = {S S S c},
        row{1} = {guard, mode=math},
        }
        \toprule
        \Gamma_3^\text{opt}/ \Gamma_1^\text{opt} & J_{3,4}^\text{opt} / \Gamma_1^\text{opt}& \Gamma_4^\text{opt} / \Gamma_1^\text{opt}& \mathcal{C}(\gopt)\\
        \midrule
        4.342962293221 & 4.871649743552 & 5.234439004767   & \num{0}   \\
        4.342962293250 & 4.871649743586 & 5.234439004803    & \num{0}   \\
        4.342962293250 & 4.871649743586 & 5.234439004803   & \num{0}   \\
        4.342962293250 & 4.871649743586 & 5.234439004803   & \num{0}   \\
        \bottomrule
    \end{tblr}
\end{table*}

Figure \ref{fig:paramsvstmax_noconsec} shows the ratios of the optimal parameters vs. \TMAX, for the BW-like Ansatz \BW.
Here, each run was initialized with the parameters $\bm{g}_\text{init} = (6 \,	12 \, 15 \,	17)^T$.
The cost function is evaluated with the 
Tanh-sinh quadrature.
It can be seen that the optimal parameters exhibit oscillations and are not converged for the displayed \TMAX. This observation is a fundamental consequence of the finite value for the cost function \copt, as the system still shows dynamics on the order of the cost function value. The parameters will not converge, if the variational Ansatz captures the EH only poorly.
Thus \TMAX will always have an influence on the optimal parameters so that no convergence will be reached if the BW-like Ansatz is used.
Additionally we see for very small \TMAX a jump in the parameters. This is an artificial solution, that occurs if the typical oscillation of the cost function can not be captured by the time window anymore.

\subsection{Convergence properties of the BW-violating Ansatz}
\label{sec:conv_bwv}
The previous discussions on convergence were based on the BW-like Ansatz \BW, where the
minimum of the cost function value was finite.
In this subsection we investigate the convergence properties of the \BWV Ansatz.
We choose $\TMAX = 1$ and the initial parameters 
$\bm{g}_\text{init} = (3 \,	5 \, 8 \,	10 \,	12 \,	14 \,	15)^T$ were used for all runs. Note that the index $i$ of a parameter $g_i$ is not directly related to the $i$-th lattice site for the Ansatz \BWV.

Table \ref{tab:midpoint_BWV_dt} shows the optimal parameters 
normalized to $\Gamma_1^\text{opt}$ for all different time steps \dt.
It can be seen that all normalized parameters agree up to twelve decimal places.
The second study concerns the influence of the maximum observation time \TMAX.
Here, the Tanh-sinh quadrature is used and each run uses a different \TMAX.
Each run is initialized with the same initial parameters as in the first study.
Table \ref{tab:midpoint_BWV_Tmax} shows the optimal parameters 
normalized to $\Gamma_1^\text{opt}$ for all different \TMAX.
All normalized parameters agree up to 13 decimal places, except for $\TMAX = 0.1$,
where only nine decimal places of the optimal ratios agree with the optimal ratios of the runs with higher \TMAX.\\
Both results demonstrate a robust behaviour with respect to the 
number of observation times and the maximum observation time.\\
The results are converged for all \dt and almost all \TMAX if a good Ansatz is chosen, i.e. an Ansatz that leads to a cost function that is numerically zero at its minimum, which is the case for the BW-violating Ansatz for the TFIM with $\Gamma = 1$. 
The mismatch between the convergence properties of the \BW and the \BWV Ansatz can be explained by the values of the cost function.
If the cost function reaches numerically zero for an Ansatz, the subsystem becomes constant in time and no dynamics is present at the optimal solution $\gopt$.
In this case $\Delta t$ and \TMAX become irrelevant.
On the other hand, if an Ansatz does not capture the EH accurately, 
the cost function will be finite at the minimum.
Thus, even at the minimum, dynamics will be present s.t. it is not irrelevant how often or how long the system is sampled.

Since it is not known a priori how accurate an Ansatz is, we conclude that it is best practice to evaluate the cost function with the Tanh-sinh quadrature to reduce the number of required measurements. \\
Note, one parameter can be fixed throughout the optimization if the Ansatz is good and the cost function drops to numerically zero.
Fixing one parameter effectively fixes the \enquote{entanglement energy scale}, which has an influence on the time scale.

\section{Violation of the BW-theorem in the XXZ model}
\label{sec:BWV_XXZ}
\begin{figure*}[t]
    \centering  
    \includegraphics{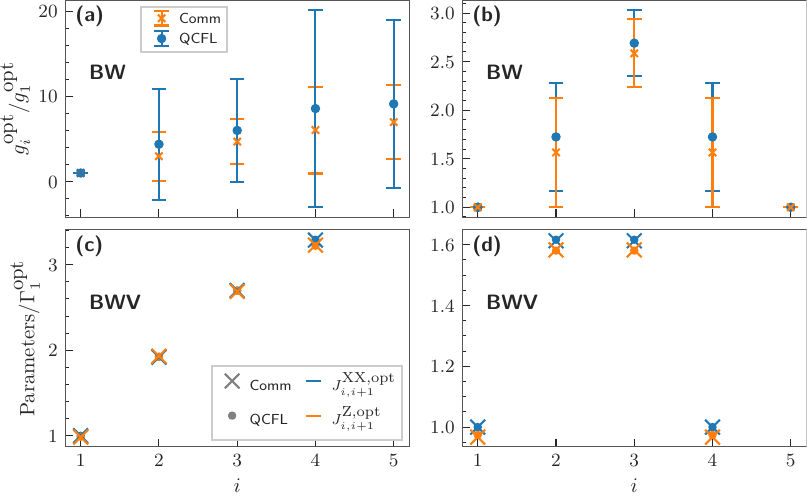}
    \caption{50 runs with different random initial parameters for the XXZ model
    with the QCFL and the commutator as a cost function (Comm).
    Outliers from the QCFL were filtered out.
    The mean value from the 50 runs are depicted.
    (a),(b) Optimal parameters for the BW-like Ansatz with OBC and 
    PBC, respectively. The errorbars are given by the standard deviation over the 50 runs (outliers from the QCFL were filtered out). 
    (c),(d) Optimal parameters for the BW-violating Ansatz with OBC and 
    PBC, respectively.
    We use $N=10$, $\NA = 5$, $\Delta = -0.5$ and open boundary conditions.}
    \label{fig:BW_vs_BWV_xxz}
\end{figure*}
Here we compare the BW-like and BW-violating Ansatz in the XXZ model with $N=10$, $\NA = 5$, $\TMAX=1$ and $\Delta = -0.5$.
The analysis is the same as for the TFIM (see Section \ref{sec:BWV_TFIM}).
The optimal parameters exhibit huge standard deviations in the case of the XXZ model with OBC and the 
BW-like Ansatz \BW (Fig. \ref{fig:BW_vs_BWV_xxz}(a)).
The differences of the optimal ratios obtained via the QCFL and the commutator show huge deviations, where as the deviations are smaller in the 
case of PBC (Fig. \ref{fig:BW_vs_BWV_xxz}(b)).
Again, the linear rise near the entanglement cut and bending at the right border become apparent in the case of OBC, while 
the symmetric behaviour can be observed if the composite system obeys PBC. 
However, the large error bars indicate that there many suboptimal or local minima, which 
are found by the optimizer.
The minimum of the cost function is not numerically zero, namely $\copt \approx \num{3.54(10)e-7}$ (OBC) and $\copt \approx \num{4.35(4)e-6}$ (PBC).
Figure \ref{fig:BW_vs_BWV_xxz}(c),(d) show the optimal parameters after filtering
obtained with the Ansatz \BWV for OBC and PBC, respectively.
The parameters obtained with the QCFL and the commutator show good agreement and exhibit very small standard deviations (see Fig. \ref{fig:BW_vs_BWV_xxz}(c),(d)).
Remarkably, in the case of OBC, the parameters only deviate slightly from the BW theorem.
That is, the ratios of the coupling in the $x$- and $y$-direction, and the coupling in $z$-direction, $\sfrac{J_{i,i+1}^\text{XX,opt}}{J_{i,i+1}^\text{Z,opt}}$,
is $1.0219 \pm 0.0005$ at most, which almost agrees with the BW theorem, predicting a value of $1$.
The deviation of the parameters from the BW theorem in the case of PBC is similar.
The cost function value at the optimum takes the value of 
$\copt \approx \num{1.12(27)e-9}$ (OBC) and $\copt \approx \num{2.51(55)e-8}$ (PBC).
The deviation from the BW theorem is not negligible, underlined by the 
drop in the cost function value by approximately three (OBC) and two (PBC) orders of magnitude if the BW-violating Ansatz is used.
Although the cost function is noticeably lower with the Ansatz \BWV
and parameters obtained via the QCFL and the commutator match very well, 
the minimum of the cost function is still finite.
\section{Methods}
\label{sec:methods}
We use the Julia programming language\cite{bezanson2017julia}.
For the optimization of the cost function, we chose the implementation of the LBFGS algorithm from the package \texttt{Optim.jl}\cite{mogensen2018optim}.
We compute the time evolution operator by matrix exponentiation.
The gradient is computed with an exact expression, utilizing the Fr\'{e}chet derivative of the time evolution operator, i.e. an matrix exponential.
For its computation we adapted the implementation from the package \texttt{ChainRules.jl}\cite{ChainRules}.
We chose the infinity norm of the gradient as a convergence criterium, which is referred to as $\nabla_\text{tol}$.
In order to stop a minimization, $\nabla_\text{tol} \leq \num{e-16}$ must hold if not mentioned otherwise.
The monitored observables are $\{ Z_i Z_{i+1} | 1 \leq i < \NA \} $.
To obtain the ground state, we use exact diagonalization if the composite system Hilbert space dimension is $ \dim \mathcal H \leq 1024$ and the Lanczos algorithm 
from the package \texttt{KrylovKit.jl}\cite{Krylovkit} otherwise.

\section{Numerical integration methods}
\label{sec:integration}
Here we briefly introduce the used integration methods.
Note that the integrand changes with each new parameters $\bm{g}$, and thus, the integral can be 
seen as a blackbox. This is why an adaptive or iterative 
integration method is a good choice to achieve convergence.
Additionally, the integrand is oscillatory, which poses another challenge.
\subsection{Right point rule}
With the right point rule, the integral over the interval $[a,b]$ is evaluated at $n$ points, equidistantly separated by a step size $\dt = \sfrac{(b-a)}{n}$ starting at $a+\dt$
\begin{equation}
    I = \int_a^b f(t)dt \approx \sum_{i=1}^{n}  f(a+i\dt) \dt.
\end{equation}
The right point rule converges linearly in $n$.
\subsection{Mid point rule}
The mid point rule is, like the right point rule, a rectangular integration method with the difference that the evaluation points are shifted to the left by $\sfrac{\dt}{2}$, s.t. the approximation reads 
\begin{equation}
    I = \int_a^b f(t)dt \approx \sum_{i=1}^{n}  f \left (a+ \left (i-\frac{1}{2} \right )\dt \right )\dt.
\end{equation}
The mid point rule converges quadratically in $n$.
\subsubsection{Tanh-sinh method}
The Tanh-sinh quadrature is one quadrature formula of a whole family, the Double Exponential Formulas\cite{Takahasi_1973} (DE Formulas). \\
Starting from an integral over the interval $[-1, 1]$
\begin{equation}
    I = \int_{-1}^1 f(t) d t,
\end{equation}
the DE Formulas utilize a variable transformation $x= \Phi (u)$ mapping the boundaries to infinity, i.e., $\phi(-\infty) = -1$ and 
$\phi(\infty) = 1$.
That is, the integral reads
\begin{equation}
    I = \int_{- \infty}^{\infty} f(\Phi (u)) \Phi' (u) \diff u. \label{eqn:DEF}
\end{equation}
Applying the trapezoidal rule with a step size $h$ to Eq. \eqref{eqn:DEF} and truncating the number of evaluation points to a finite value yields
\begin{equation}
    I_h = h \sum_{j = -M}^{M} f(\Phi (jh)) \Phi' (jh)
\end{equation} 
The error stems from
discretization and the truncation of the infinite sum \cite{murota2023doubleexponentialtransformationquickreview}.
The best balance between the discretization error and the truncation error is achieved by a variable transformation for which the integrand decays double exponentially, giving the DE Formulas their name.
The double exponential decay is achieved by the variable transformation
\begin{equation}
    \Phi (u) = \tanh \left ( \frac{\pi}{2} \sinh (u) \right ), \label{eqn:tanh-sinh}
\end{equation}
giving the Tanh-sinh quadrature its name.
We adapted the implementation from the package \texttt{DoubleExponentialFormulas.jl}\cite{DEformulas} and added a reliable error estimation scheme presented in Reference \cite{em/1128371757}.
\subsection{Gauß-Legendre quadrature}
The $n$-point Gauß-Legendre quadrature approximates the integral 
\begin{equation}
    \int_{-1}^1 f(t)dt = \sum_{i=0}^{n-1} w_i f(t_i),
\end{equation}
where $t_i$ are the roots of the $n$-th Legendre polynome $P_n$ and the weights are given by\cite{johansson2018fastrigorousarbitraryprecisioncomputation}\begin{equation}
    w_i = \frac{2}{(1-t_i^2)(P'_n(t_i))^2}.
\end{equation}
We use the implementation of the package \texttt{FastGaussQuadrature.jl}\cite{fastgauss}.
\subsection{Adaptive Gauß-Kronrod quadrature}
The Gauß-Kronrod quadrature is an extension of the standard Gauß-quadratures. 
Starting from a $n'$-point Gauß-quadrature, they are extended with $n'+1$ additional points to a $2n'+1$ quadrature scheme. This has the advantage that the Gauß-Kronrod quadrature has a good error estimate by computing the difference of the standard Gauß quadrature and its Kronrod extension, where in contrast a standalone Gauß-quadrature cannot reuse already computed function values to estimate an error s.t. it would need to compute more function values in order to get an error estimate. For more details, we refer to the documentation of the used package \texttt{Quadgk.jl}\cite{quadgk} and the seminal work by Laurie\cite{laurie}.
\section{Optimizer benchmark}
\label{sec:benchmarks}
\begin{figure*}
    \centering
    \includegraphics{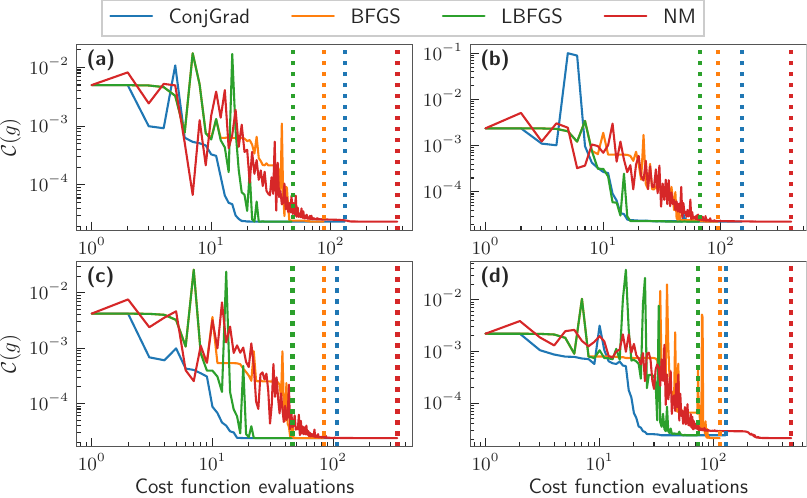}
    \caption{Cost function value vs. number of cost function evaluations in one minimization run.
    Random initial parameters are sampled and the cost function is minimized with the different algorithms.
    This procedure is done four times (run one, two, three and four in (a),(b),(c) and (d), respectively).
    The vertical dotted lines indicate the points, where the optimizers are converged.
    We use the TFIM with $N=8$, $\NA = 4$, $\Gamma = 1$, open boundary conditions and the BW-like Ansatz \BW.}
    \label{fig:benchmarks_optimizer}
\end{figure*}
This section shows that the used method for optimization is the best methods for this kind of problem 
among all the methods, which are used for the comparison.
The model used is the TFIM with $N=8$, $\NA = 4$, OBC and $\Gamma = 1$
with the variational Ansatz \BW.

The algorithms used for comparison are Conjugate Gradient (ConjGrad), BFGS and Nelder-Mead (NM),
which are all implemented in the package \texttt{Optim.jl} as well.
The maximum integration time is set to $\TMAX = 1$.
The first three algorithms are gradient-based, whereas the Nealder-Mead algorithm is a direct search method and therefore does not 
require information about the gradient or Hessian. 
Initial parameters are chosen at random and the cost function is minimized with all different algorithms.
The convergence criterium for the gradient-based methods is set to $\nabla_\text{tol} = \num{e-16}$.
NM uses a so called simplex, which consists of multiple points in the parameter space.
The convergence criterium here is the standard deviation of the cost function value at these points in the current simplex, since 
the cost function value at the points of the simplex should be equal in the vicinity of the minimum.
In this test, the standard deviation is required to be less than or equal to $\num{e-16}$, too.
The cost function value vs. the number of cost function evaluations in one minimization run is shown
in Fig. \ref{fig:benchmarks_optimizer}.
For all four different initial parameters, the LBFGS algorithm needed the fewest cost function evaluations, as indicated by 
the vertical dotted lines.
However, in the fourth run (Fig. \ref{fig:benchmarks_optimizer}(d)),
the BFGS and Nelder-Mead algorithms found a slightly lower minimum ($\approx \num{3e-5}$ smaller) than the BFGS and Conjugate Gradient algorithms.
Table \ref{tab:opttimes} lists the time it took to minimize the cost function among all runs for all different algorithms.
The minimization was repeated ten times for one set of initial parameters 
and the smallest time over these ten repetitions is listed.  
In all runs, the LBFGS algorithm won again in terms of runtime.
\begin{table}
    \centering
    \caption{Runtime of the algorithms for the four sets of random initial parameters 
    (i.e. four different runs).
    We use the TFIM with $N=8$, $\NA = 4$, $\Gamma = 1$, open boundary conditions and the BW-like Ansatz \BW.}
    \label{tab:opttimes}
    \sisetup{table-format=1.3}
    \begin{tblr}{
        colspec = {c S S S S},
        row{1} = {guard, mode=math}, row{2} = {guard, mode=math},
        }
        \toprule
        & \SetCell[c=4]{c} \text{runtime} \mathbin{/} \unit{\second}  & & & \\
        \cmidrule[lr]{2-5}
        \text{algorithm} & \text{run 1} & \text{run 2} & \text{run 3} & \text{run 4}\\
        \midrule
        \text{ConjGrad}    & 2.400 & 3.088 & 1.898 & 2.271 \\
        \text{BFGS}        & 1.727 & 1.914 & 1.625 & 2.494 \\
        \text{LBFGS}       & 0.940 & 1.372 & 0.909 & 1.377 \\
        \text{NM}          & 1.873 & 2.161 & 1.779 & 2.505 \\
        \bottomrule
    \end{tblr}
\end{table}
To conclude, even tho the LBFGS algorithm did not find the lowest minimum in the fourth run, the efficiency is very convincing.
Additionally, this happened only one out of four times with random initial parameters. 
With a good initial guess this should not happen.
The Gradient descent, ADAM and Simulated annealing algorithm have also been tested but not listed, because 
their performance were much worse than the algorithms included in the benchmark. 
It needs to be mentioned that a model with only 4 parameters has been used.
To get deeper insights into the performance of the optimization algorithms, a model with significantly more 
parameters could prove helpful.
\section{Long range corrections in the XXZ model}
\label{sec:correctionsXXZ}
\begin{table}[t]
    \centering
    \caption{Optimal parameters $\{ J_{i,i+r}^\text{XX}, J_{i,i+r}^\text{Z}\}$ including all long-range interactions
    for $\rmax = 4$ rounded to four decimal places.
    Here we use $N=10$, $N_\text{A} = 5$, $\Delta = -0.5$ and open boundary conditions.}
    \label{tab:Jvsrmax}
    \sisetup{table-format=2.4}
    \begin{tblr}{
        colspec = {c S S S S S},
        row{1} = {guard, mode=math},
        }
        \toprule
        & r = 1  & r=2 & r=3 & r=4 \\
        \midrule
        $J_{1,1+r}^\text{XX}$   &   0.8435 &    0.0056& 0.0013   & 0.0032   \\
        $J_{1,1+r}^\text{Z}$    &   0.8821 &    0.0482& 0.0447   & 0.0474   \\
        $J_{2,2+r}^\text{XX}$   &   1.6238 &   -0.0081& 0.0027   &           \\
        $J_{2,2+r}^\text{Z}$    &   1.6379 &   0.0111 & 0.0265   &           \\
        $J_{3,3+r}^\text{XX}$   &   2.3228 &   0.0673 &           &           \\
        $J_{3,3+r}^\text{Z}$    &   2.3268 &   0.0975 &           &           \\
        $J_{4,4+r}^\text{XX}$   &   2.8433 &           &           &           \\
        $J_{4,4+r}^\text{Z}$    &   2.6806 &           &           &           \\
        \bottomrule
    \end{tblr}
\end{table}

\begin{figure}[t]
    \centering
    \includegraphics{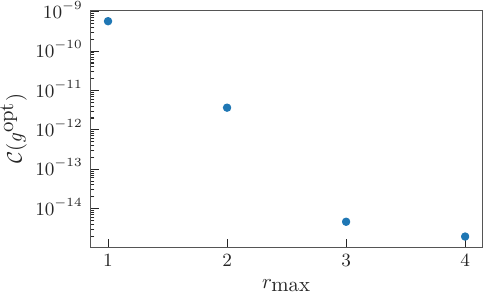}
    \caption{Cost function $\mathcal{C}(\gopt)$ at the found solution vs. the maximum range of interaction $r_\text{max}$ included
    in the correction term.
    Here we use $N=10$, $N_\text{A} = 5$, $\Delta = -0.5$ and open boundary conditions.}
    \label{fig:costforrmax}
\end{figure}

We investigate if the cost function for the XXZ can be further reduced by including long-range corrections to the variational entanglement Hamiltonian. 
The variational Ansatz, together with the corrections $\corr$, reads
\begin{equation}
    \varwo = \BWV + \corr \\ 
    = \sum_{r=1}^{r_\text{max}} \sum_{i=1}^{N_A-r} \left ( J^\text{XX}_{i, i+r}  \left ( X_i X_{i+r} + Y_i Y_{i+r} \right ) 
    +  J^\text{Z}_{i, i+r} \increment  Z_i Z_{i+r} \right ),
\end{equation} 
where $\{J_{i,i+r}^\text{XX}, J_{i,i+r}^\text{Z}\}$ act as variational parameters.
The quantity $r_\text{max}$ determines the maximum range of interaction.
Every term beyond $r=1$ is a part of the long-range interactions, and thus, part of the corrections.
The optimal parameters from Section \ref{sec:BWV_XXZ} of the BW-violating Ansatz \BWV are used for initialization of the parameters 
for $r > 1$. 
All long-range couplings (beyond $r=1$) are initialized to zero, as these are expected to be small.  
Table \ref{tab:Jvsrmax} lists the optimal parameters for a run with $r_\text{max} = 4$, i.e.,
all long-range terms included.
It can be seen that the corrections are at least one magnitude smaller in comparison  to the parameters for $r=1$.
The parameters show a decay with the range of interaction $r$. 
We observe in Fig. \ref{fig:costforrmax}, that the variational Ansatz containing long-range corrections indeed reduces the value of the cost function, demonstrating that such corrections to the BW ansatz are crucial to represent the EH.
Note that the norm of the gradient was very low in case of $\rmax = 3$ and $\rmax = 4$, and we terminated the optimization after $100000$ iterations.

\section{Extrapolation into the thermodynamic limit}
\label{sec:extr}
\begin{figure}
    \centering
    \includegraphics{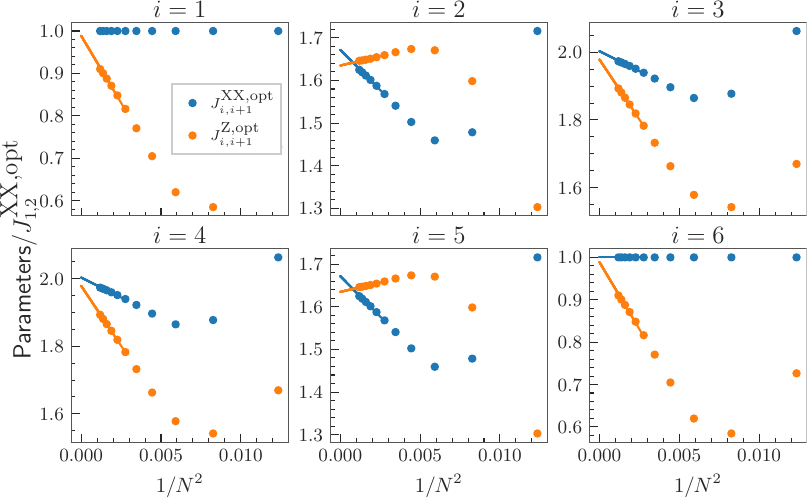}
    \caption{Optimal parameters $J_{i,i+1}^\text{XX,opt}$ and $J_{i,i+1}^\text{Z,opt}$ normalized to 
    $J_{1,2}^\text{XX,opt}$ vs. $\sfrac{1}{N^2}$ for each lattice site $i$ in the subsystem A for $\NA = 7$, $\Delta = -0.5$ and PBC. 
    The solid lines are linear fits.}
    \label{fig:extrap_optim}
\end{figure}

To obtain the couplings $J_{i,i+1}^\text{XX,opt}$ and $J_{i,i+1}^\text{Z,opt}$ in the TDL,
the procedure is the following:
\begin{enumerate}
    \item Extract the ground state of the system Hamiltonian for different lattice sizes $N$ (up to 
    $N = 29$ could have been achieved with the Lancos algorithm).
    \item Construct the RDM with the ground state obtained in step one for a subsystem chain length \NA.
    \item Run the algorithm with the BW-violating Ansatz \BWV 
    with  $J_{i,i+1}^\text{XX}$ and $J_{i,i+1}^\text{Z}$ as variational parameters for the different RDMs for each composite system size $N$
    from step 2 for a subsystem chain length \NA.
    \item Plot the ratios of the obtained parameters $\sfrac{J_{i,i+1}^\text{XX,opt}}{J_{1,2}^\text{XX,opt}}$ and $\sfrac{J_{i,i+1}^\text{Z,opt}}{J_{1,2}^\text{XX,opt}}$ vs. $\sfrac{1}{N^2}$ and extrapolate for $\sfrac{1}{N} \to 0$, i.e., into the TDL.
    \item Repeat step two to four for different subsystem chain lengths \NA.
\end{enumerate}
Figure \ref{fig:extrap_optim} shows the obtained optimal parameters normalized to $J_{1,2}^\text{XX,opt}$ vs. $\sfrac{1}{N^2}$ for $\NA = 7$.
The index $i$ indicates the lattice site and the solid lines are the corresponding fits.
For brevity, the plots for other \NA are omitted.
A linear fit was used
\begin{equation}
    \frac{J_{i,i+1}^{\Omega,\text{opt}}}{J_{1,2}^\text{XX,opt}} \left (N \right ) = p_1 \frac{1}{N^2}+ p_2 , \quad \Omega = \text{XX}, \text{Z} \label{eqn:fitodd}
\end{equation}
where $p_1$ and $p_2$ act as parameters for the fit.
The optimal parameters normalized to $J_{1,2}^\text{XX,opt}$ extrapolated in to the TDL ($N \to \infty$) then are 
\begin{equation}
    \frac{J_{i,i+1}^{\Omega,\text{opt}}}{J_{1,2}^\text{XX,opt}} (N \to \infty) = p_2 , \quad \Omega = \text{XX}, \text{Z}.
\end{equation} 
\section{Zero-noise extrapolation}
\label{sec:ZNE}
For each point in the phase diagram and number of measurements $N_\text{M} \in [10^3,10^9]$ (TFIM) and $N_\text{M} \in [10^3,10^{11}]$ (XXZ model), we sampled the cost function ten times at the optimal parameters obtained with optimizations without noise and averaged over these ten samples.
We used the Gauß-Legendre quadrature with five time points and $T_\text{max} = 2$.
Our fit model is based on the fact that the cost function is a second order polynomial in $\sfrac{1}{\sqrt{N_\text{M}}}$ in the presence of noise
\begin{equation}
    C(N_\text{M}; \bm{g}_\text{opt}) = a \frac{1}{N_\text{M}} + b \frac{1}{\sqrt{N_\text{M}}} + c.
\end{equation}
Since we want to fit over several orders of magnitude, we first need to take the logarithm of our data to prevent that the higher data points dominate the errors in the least-square algorithm.
Our final fit model $y(x)$ is
\begin{equation}
    y(x) = \text{log}_{10}(a 10^{2x} + b 10^x + c) 
\end{equation}
with $x = \text{log}_{10}(\sfrac{1}{\sqrt{N_\text{M}}})$.
The extrapolated value, i.e. $N_\text{M} \to \infty$,  then is the fit parameter $c$.
\section{XXZ model across the phase diagram for larger systems}
\label{sec:xxz_phase}
\begin{figure}[h]
    \centering
    \includegraphics{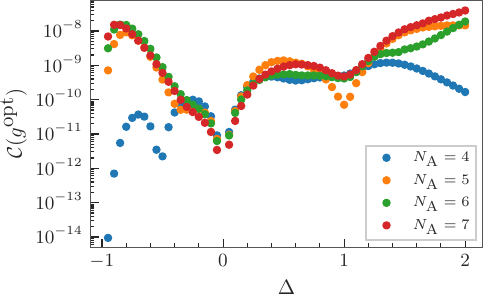}
    \caption{Minimum of the cost function for the XXZ model with varying anisotropy $\Delta$ for different system sizes.
    $N=2\NA$ with OBC always holds and we use the BW-violating Ansatz \BWV.}
    \label{fig:exXXZallN}
\end{figure}

Here, in Fig. \ref{fig:exXXZallN}, we show the minimum of the cost function in dependence on the anisotropy $\Delta$ for different system sizes in the XXZ model without noise. 
$N=2\NA$ with OBC always holds and we use the BW-violating Ansatz.
All values below double precision have been omitted.
\end{document}